\documentclass[useAMS,usenatbib]{mn2e}

\usepackage{graphicx,amssymb,bm}
\usepackage{float, Abbrev}
\usepackage[fleqn]{amsmath}  
\usepackage{color}

\newcommand{\be}{\begin{equation}}
\newcommand{\ee}{\end{equation}}
\newcommand{\ba}{\begin{eqnarray}}
\newcommand{\ea}{\end{eqnarray}}

\newcommand{\sdm}{\sigma}
\newcommand{\mdm}{m}
\newcommand{\sstar}{\sigma^\star}
\newcommand{\rmd}{\mathrm{d}}
\newcommand{\Vs}{_\mathrm{S}}
\newcommand{\Vg}{_\mathrm{G}}
\newcommand{\Vd}{_\mathrm{D}}

\newcommand{\bv}[1]{\mbox{\boldmath$#1$}}

\def\simlt{\lower.5ex\hbox{$\; \buildrel < \over \sim \;$}}
\def\simgt{\lower.5ex\hbox{$\; \buildrel > \over \sim \;$}}

\usepackage{graphicx} 

\title[On the cross-section of Dark Matter using substructure in-fall]{On the cross-section of Dark Matter using substructure infall into galaxy clusters}
\author[D.\ R.\ Harvey et al.]{David Harvey$^{1}$\thanks{e-mail: {\tt drh@roe.ac.uk}}, Eric Tittley$^{1}$, Richard Massey$^{2}$, Thomas D.\ Kitching$^{3}$, Andy Taylor$^{1}$, \newauthor Simon R. Pike$^{4}$, Scott T. Kay$^{4}$, Erwin T.\ Lau$^{5,6}$ and Daisuke Nagai$^{5,6}$\\
$^{1}$SUPA, University of Edinburgh, Royal Observatory, Blackford Hill, Edinburgh EH9 3HJ, UK \\
$^{2}$Institute for Computational Cosmology, Durham University, South Road, Durham DH1 3LE, UK \\
$^{3}$Mullard Space Science Laboratory, University College London, Holmbury St Mary, Dorking, Surrey RH5 6NT, UK\\
$^{4}$Jodrell Bank Centre for Astrophysics, School of Physics and Astronomy, The University of Manchester, Manchester, M13 9PL, UK\\
$^{5}$Department of Physics, Yale University, New Haven, CT 06520, USA\\
$^{6}$Yale Center for Astronomy \& Astrophysics, New Haven, CT 06520, USA}
\begin{document}

\date{Accepted ---. Received ---; in original form \today.}

\pagerange{\pageref{firstpage}--\pageref{lastpage}} \pubyear{2013}

\maketitle

\label{firstpage}

\begin{abstract}
\noindent 
We develop a statistical method 
to measure the interaction cross-section of Dark Matter, 
exploiting the continuous {\it minor merger} events in which small substructures fall into galaxy clusters.
We find that by taking the ratio of the distances between the galaxies
 and Dark Matter, and galaxies and gas in accreting sub-halos, we form a quantity that can be
 statistically averaged over a large sample of systems whilst  removing
 any inherent line-of-sight projections.
In order to interpret this ratio as a cross-section of Dark Matter we derive an analytical description of sub-halo infall which encompasses; the force of the main cluster potential, the drag on a gas sub-halo, a model for Dark Matter self-interactions and the resulting sub-halo drag, the force on the gas and galaxies due to the Dark Matter sub-halo potential, and finally the buoyancy on the gas and Dark Matter. 
We create mock observations from cosmological simulations of structure formation and find that collisionless Dark Matter  becomes physically separated from X-ray gas by up to $\sim20h^{-1}$~kpc.
Adding realistic levels of noise, we are able to predict achievable constraints from observational data.
Current archival data should be able to detect a difference in the dynamical behaviour of Dark Matter and standard model particles at 6$\sigma$, and measure the total interaction cross-section $\sdm/m$ with 68\% confidence limits of $\pm1\,{\rm cm}^2{\rm g}^{-1} $.
We note that this method is not restricted by the limited number of major merging events and is easily extended to large samples of clusters from future surveys which could potentially push statistical errors to $<0.1$cm$^2$g$^{-1}$.

\end{abstract}


\section{Introduction} \label{sec:intro}
The prevailing cosmological model indicates that 85\% of the mass content of the Universe is Dark Matter (DM) that theoretically couples only weakly (or not at all) to baryons through the electroweak force.
Because of the enormous practical difficulty of detecting DM in laboratories, remarkably little is known about its properties.
Deciphering its nature remains one of the most outstanding questions in physics \citep{PeterRev}. 

DM does interact via gravity, and its abundance means that it dominates the gravitational mass on scales $>1$kpc.
However, theoretical predictions from models of non-interacting, cold Dark Matter \citep{satelliteprob} overpredict the observed abundance and central concentration of substructure on small ($\simlt 100$~kpc) scales \citep{corecusp}.
This discrepancy can be resolved if DM has a finite interaction cross-section with itself or standard model particles \citep{ObserveSIDM, GalaxySIDM} --- and weak coupling is a generic consequence in several extensions to the Standard Model, \citep[e.g.][]{ SecludedDM,ChargedDM, DMYukawa}.

A decade ago, self-interacting DM was thought to be ruled out by negative results on tests for sphericity \citep{SIDMTest}, cores \citep{HaloSIDM, SIDMCore}, and sub-halo evaporation \citep{EllGalSIDM} in galaxy clusters. 
However, recent high resolution simulations show that self-interactions produce much more triaxial inner halos \citep{SIDMSim}, smaller cores \citep{SIDMSimA, SubhalosSIDM}, and less evaporation than previously thought.
A self-interaction cross-section per particle mass, $\sigma/m\approx1$~cm$^2/$g remains as consistent with observations as non-interacting CDM. 

The largest bound structures in the Universe are galaxy clusters which are collections of several thousands of galaxies, each surrounded by vast ($>10^{14}M_\odot$) quantities of DM and mainly ionised hydrogen gas.
The highly-successful cold Dark Matter model of structure formation predicts that galaxy clusters grow hierarchically, by continually accreting smaller groups of galaxies and occasionally colliding.
Such minor and major merging events offer a unique laboratory in which to investigate the particle physics of DM.
Compared to terrestrial colliders, the energy per particle during a merger is small ( a factor of $10^{-6}$ less than that at LHC), but the cumulative number density of Dark Matter particles is enormous with collisions involving up to  $\sim$$10^{70}$ particles per major merging event  (assuming $\mdm=10$~GeV Dark Matter particles).


It is possible to map the locations of all components of a galaxy cluster.
Intracluster gas in galaxy clusters emits bremstrahlung radiation, which is visible at X-ray wavelengths \citep{bremm}, whilst the
DM component can be mapped via gravitational lensing \citep{BS01,RefregierRev,HoekstraRev,MKRev}.
Several studies of individual clusters have constrained $\sdm/m$ by observing the separation of DM from gas in the aftermath of a collision leading to constraints on the total interaction cross-section per unit mass of DM:
1ES 0657-558 \citep{bulletclusterA,separation,bulletclusterB}; MACSJ0025.4-1222 \citep{minibullet}; 
A520 \citep{A520,A520A,A520B}; A2744 \citep{A2744}; DLSCL J0916.2+2951 \citep{musket}.
Each cluster constitutes three components: the member galaxies, the intracluster baryonic gas, and the DM halo. 
The components' different interaction cross-sections make them behave differently during the collision.
Galaxies act as collisionless test particles, passing through the collision unimpeded (except via gravity).
The large cross-section of baryonic gas makes it lag behind the galaxies.
Non-interacting DM should remain with the galaxies, and interacting Dark Matter should lie between the galaxies and the gas (tending to the position of the gas as $\sdm$ tends to the effective cross section of Hydrogen).

Unfortunately, major merger events, observed shortly after first core passage for maximal observed separation of components, are rare in the Universe \citep{rareevents, extremeObjects}.
Constraints on $\sdm$ from a small number of systems are fundamentally limited by their unknown impact velocity, impact parameter and angle with respect to the line of sight \citep{bulletcluster,impactpars}.


\begin{figure}
	\centering
	\label{fig:radial}\includegraphics[width=84mm]{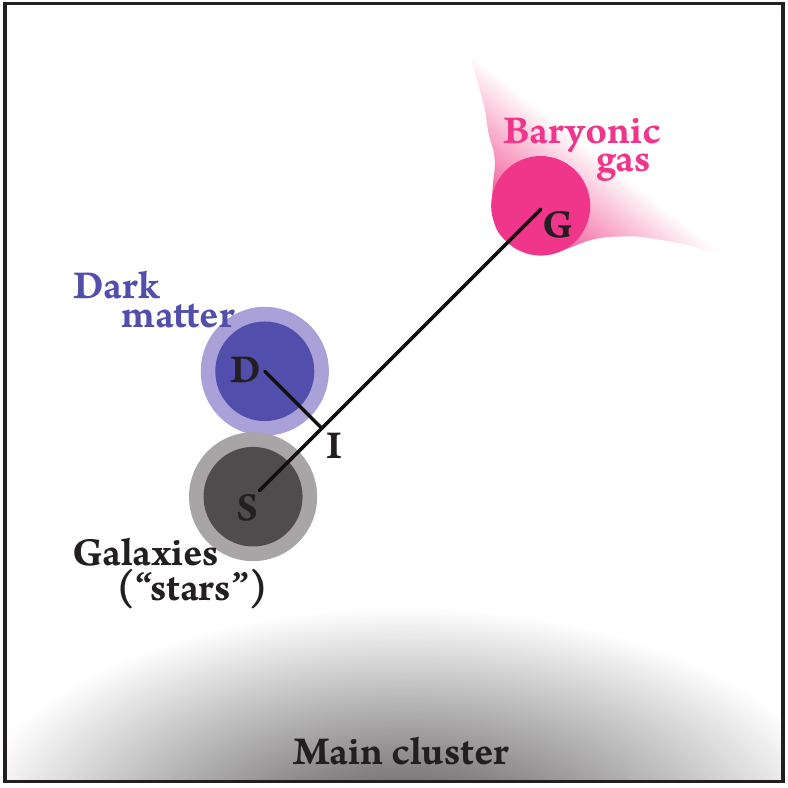}
\caption{Cartoon illustrating how we propose to use all three components of infalling substructure. 
The vector from galaxies to gas defines an (approximate) direction of infall.
Dark matter should lie some fraction along this vector, depending upon its interaction cross section. 
The observed positions will be noisy, so in practice we will measure the parallel and perpendicular vectors from galaxies to Dark Matter.
If $\sdm$$=0$, these should both average to zero.
Throughout this paper we adopt shorthand subscript notation G for gas, D for Dark Matter, I for the intersection point closest to the Dark Matter in the direction towards the gas, and S (``stars'') for galaxies.
\label{fig:system}}
\end{figure}

As suggested by \citet[][hereafter MKN11]{bulleticity}, the separation between galaxies, gas and DM can also be measured in minor mergers.
The displacement of gas and Dark Matter from galaxies is likely to be much smaller than in major mergers.
However, minor merger events are the dominant growth mechanism for large-scale structure in the Universe, and most clusters are accreting a piece of substructure around $\sim$$10\%$ of their total mass at any time \citep[e.g.][]{Powell09}. 
Analysis of a sufficiently large observed sample of minor mergers should yield much tighter constrains on $\sdm/m$ than a small number of major mergers, while automatically averaging over systematic uncertainty in orbital parameters.
The statistically averaged offset stacked over many pairs of DM and gas sub-halos was coined ``bulleticity'', and can be obtained from potentially hundreds of thousands of clusters across the sky.

Using hydrodynamical simulations of ordinary clusters, MKN11 found that substructures' DM and gas components become separated during infall by $|\bv{b}|\simeq 10\arcsec$ ($18~h^{-1}$~kpc) at $z=0.1$.
Observing such small separations requires high precision DM astrometry.
This is easily achievable using strong gravitational lensing. 
Indeed, \citet[][hereafter WS11]{cannibal} discovered a $\sim$$3\arcsec$ offset between the DM {\it and the sub-halo galaxies} in A3827 (at $z=0.1$) that implies a tantalising first detection of weakly interacting DM. 
However, a statistical bulleticity measurement relies upon measurements from a very large sample of clusters, and strong lensing of substructure is rare.
\cite{Harvey13} therefore showed that with current data one would be able to use weak gravitational lensing to constrain the positions of substructure which could then be applied to all clusters.
Using a parametric mass map reconstruction and marginalising over ``nuisance'' parameters that here include cluster mass and concentration, they achieved a precision on the position of simulated clusters' main- and sub-halos of better than $1\arcsec$. 

The main limitation of this technique is that substructures are observed both falling into the cluster and heading out. 
DM lies closer to the cluster centre than gas during infall, but the situation is reversed after core passage.
Measuring absolute offsets is difficult because the signal in individual clusters' mass maps can be confused with shot noise (for {\sl HST} observations of a single cluster, positional accuracy is limited to $\sim$$10\arcsec$, \citealt{Harvey13}).
Since most observable substructures at high redshift are still falling in to a cluster, MKN11 suggested separating the offsets into radial and tangential components.
In principle, this permits a statistically robust measurement of the radial separation $b_r$, in which DM is closer to the cluster centre than the gas and noise averages to zero.
It also permits a simultaneous null test, because symmetry requires the mean signal (and noise) of the tangential separation $b_t$ to also be zero.
Unfortunately, MKN11 found in simulations that the radial bulleticity signal 
is an order of magnitude smaller than the absolute bulleticity at $z=0.6$, and it becomes vanishingly small at $z=0$.
Measuring this signal would be observationally challenging, and interpreting it may rely upon accurate cosmological simulations that specify the merger history.



Extending the idea laid out by MKN11 we develop a statistical technique for measuring $\sdm/m$ from a large sample of major and minor mergers.
Building on the earlier idea of averaging over many collision scenarios, this new method breaks previous degeneracies by using the galaxy component to define the motion of the sub-halo, and the ratio of the distances from the DM and gas component to the galaxy component to remove uncertainties in the projection orientation to our line of sight. By using the distance from the gas to the galaxies we will be able to calibrate any finite offset between the DM and the galaxies resulting in a cross-section measured directly from data.

This paper is organised as follows. 
In Section \ref{sec:cross} we develop an analytic model of substructure infall into a cluster, which we can use to develop qualitative understanding of the effects of DM interactions, and to quantitatively interpret future observations.
In Section \ref{sec:results} we apply our method on mock data from full hydrodynamical simulations of galaxy clusters embedded in the standard cosmological model.
In Section \ref{sec:data} we estimate expected constraints on various parameters from realistic data.
We discuss our results and conclude in Section \ref{sec:conc}.

\section{Methodology}\label{sec:cross}

Here we present a new method to constrain $\sdm/m$ from minor mergers. 
We exploit the fact that each piece of substructure contains {\it three} components (galaxies, gas and Dark Matter), from which {\it two} 2D offsets can be measured independently.
By measuring the ratio of the observed offset between the galaxies and Dark Matter and the offset between the galaxies and the X-ray gas, one can consider a parameter which is independent of any projection degeneracies.
In order to interpret this parameter for a measurement of $\sdm/m$, we derive an analytical prescription of sub-halo infall including all relevant forces such as; the cluster potential, the DM sub-halo potential, drag on the gas, DM interactions and the resultant drag on a DM halo and buoyancy.

As illustrated in Figure~\ref{fig:system}, we incorporate all information of the sub-halo system into our analysis. 
Compared to MKN11, the two extra pieces of information define (i) a new preferred direction and (ii) a calibrated scale length.
We shall probe the cross-section through the offset between the galaxies and DM, but interpret it in terms of the offset between the galaxies and the gas.
Throughout this paper we adopt shorthand subscript notation G for gas, D for Dark Matter, I for the intersection point closest to the Dark Matter in the direction towards the gas, and S (stars) for galaxies.
\subsection{Calibrating $\sdm/m$ with relative distances}

We assume that substructure member galaxies act as collisionless test particles during infall, acted upon by only the force of gravity.
We also assume that the main extra force acting on the baryonic gas is a drag force from the intracluster medium (ICM), which gradually separates it from the galaxies.
Crucially, this offset defines a unique displacement vector $\bv{d}_{\mathrm{SG}}=\overline{SG}$ that is antiparallel to the direction of motion, whether the substructure is falling into or emerging out of the cluster.
We propose measuring the position of the Dark Matter with respect to this direction.

The observed position of DM will depend upon its interaction cross-section.
If $\sdm=0$, the collisionless DM will remain with the galaxies.
If $\sdm>0$, forces on the DM will be exerted in the same direction\footnote{The substructure's DM could potentially interact with both the cluster ICM (DM-baryon interactions) and the cluster DM (DM-DM interactions).
These cluster components will have slightly different physical extent.
If DM-baryon interactions dominate, the substructure DM will experience a force in the same direction and at the same time as the substructure gas.
If DM-DM interactions dominate, the force could start acting earlier and in a slightly different direction, but we shall neglect this for now.} as those on the gas, and it will move some fraction $\bv{d}_{\mathrm{SI}}$ ($\overline{SI}$ in Figure \ref{fig:system}) along the vector.
Using the galaxies to define the direction of motion ensures that this should remain positive even if the substructure has already passed through the main cluster, thus preserving our signal rather than averaging most of it away.
Furthermore, symmetry again provides a null test.
Regardless of the origin and nature of the forces, a lack of preference for apparently clockwise or anticlockwise mergers still demands that the mean observed perpendicular offset of Dark Matter from the infall direction, $\langle\bv{d}_{\mathrm{DI}}\rangle$, ($\overline{DI}$ in Figure \ref{fig:system}) must be consistent with zero.





We propose calibrating the observed offset of substructure DM against the offset of substructure gas, whose properties are well known and understood.
We form the ratio
\be
\label{eqn:alpha}
\beta\equiv\frac{|\bv{d}_{\mathrm{SD}}|}{|\bv{d}_{\mathrm{SG}}|}=\frac{{d}_{\mathrm{SD}}}{{d}_{\mathrm{SG}}}.
\ee
The cross-section of Dark Matter should be measurable from 
\be
\label{eqn:alphaparallel}
\beta_ \parallel\equiv\frac{{d}_{\mathrm{SI}}}{{d}_{\mathrm{SG}}},
\ee
where we choose a Greek rather than Roman letter to denote the dimensionless quantity.
The simultaneous null test for systematics can be measured through
\be
\label{eqn:alphaperpendicular}
\beta_ \perp\equiv\frac{{d}_{\mathrm{DI}}}{{d}_{\mathrm{SG}}},
\ee
which should be consistent with zero in a  large sample.
Any deviation from this should reflect the statistical error in the positional estimates of DM.

Introducing a ratio has advantages and disadvantages.
The great advantage of taking this ratio, is that every individual measurement is now invariant to changes in the orientation of the merger with respect to the line of sight.
If the merger is viewed in the plane of the sky, all the apparent angular distances will be large, and the signal-to-noise ratio will be maximised.
If the merger occurs close to the line of sight, the apparent angular distances in both the numerator and denominator will shrink equally: the signal will remain the same, but will be measured with more noise. 
This makes it possible to combine the observed values of $\langle\beta_\parallel\rangle$ and $\langle\beta_\perp\rangle$ from a large sample of bullets via a simple weighted mean.
One disadvantage is that noise on both a numerator and denominator can lead to non-Gaussian or even biased error distributions, which we will need to treat with care.

\subsection{A physical model of Dark Matter and gas infall}

The accretion of substructure onto a cluster is a complex process that requires sophisticated hydrodynamical simulations to model completely.
However, we can build an approximate analytic model that will aid understanding and should be sufficiently accurate to interpret an initial detection of $\beta$.
Notably, we shall add sufficient complexity to deal with many of the known limitations of previous work.
 

Let us first explicitly define the forces acting on the three components, galaxies, gas, DM, of substructure falling into a cluster.
Following MKN11, we assume the distribution of mass in the cluster is a singular isothermal sphere with characteristic density $\rho_0 $ at radius $r_0$, although as we shall we, the precise form does not matter.
In addition to gravitational attraction towards the cluster, the gas will feel a drag force, $\bv{D}\Vg$, and the DM a drag force, $\bv{D}\Vd$, plus gas and DM will feel a buoyancy, $\bv{B}\Vg$ and $\bv{B}\Vd$ respectively, due to particle-particle interactions within the ICM. 
There is also a gravitational attraction of the galaxies and gas towards the substructure's dominant DM component, $\bv{G}_\mathrm{SD}$ and $\bv{G}_\mathrm{GD}$. We neglect the gravitational influence of the other, less massive components.
In the reference frame of the cluster the equations of motion for the substructure galaxies, gas and Dark Matter, are respectively,
\be
\frac{\rm d^2\bv{r}\Vs}{\rm d t^2} =
-\frac{4\pi G\rho_0r_0^2}{r\Vs^2}\bv{r}\Vs+\frac{\bv{G}_\mathrm{SD}}{M\Vs},
\label{eqn:star_acc}
\ee
\be
\frac{\rm d^2\bv{r}\Vg}{\rmd t^2} =
-\frac{4\pi G\rho_0r_0^2}{r\Vg^2}\bv{r}\Vg+\frac{\bv{D}\Vg}{M\Vg}+\frac{\bv{B}\Vg}{M\Vg}+\frac{\bv{G}_\mathrm{GD}}{M\Vg}
\label{eqn:gas_acc}
\ee
\be
\frac{\rmd^2\bv{r}\Vd}{\rmd t^2} =
-\frac{4\pi G\rho_0r_0^2}{r\Vd^2}\bv{r}\Vd+\frac{\bv{D}\Vd}{M\Vd}+\frac{\bv{B}\Vd}{M\Vd},
\label{eqn:dm_acc}
\ee
where $M_S$, $M_G$, $M_D$ are the masses of the galaxy, gas and DM component respectively.
\begin{figure}
		\begin{centering}
			\includegraphics[width = 8.5cm]{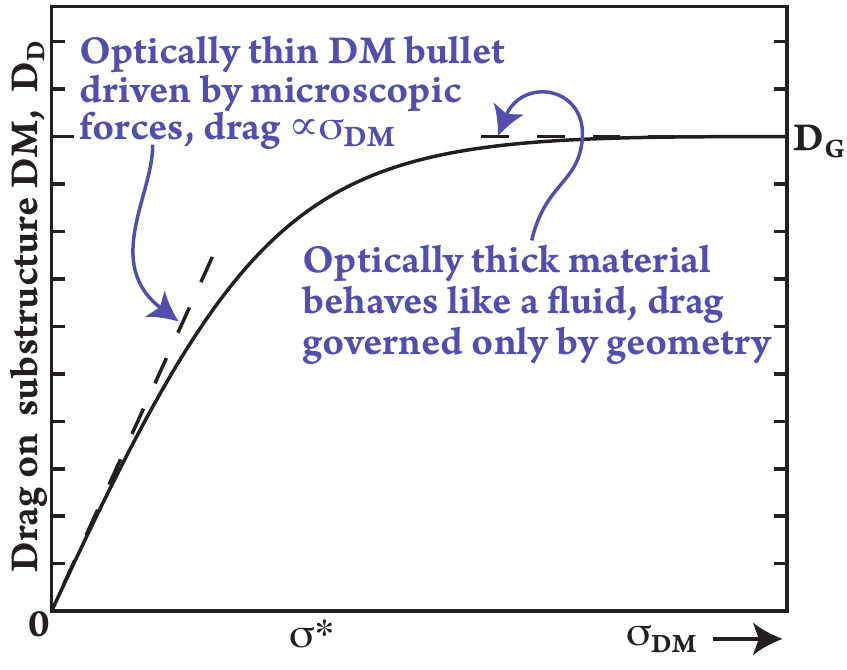}
			\caption{\label{fig:beta}
			Generic behaviour of drag force acting on Dark Matter substructure, as a function of interaction cross-section.
			We propose an interpolation function between the two well-understood extremes based on optical depth.
			This function is essential to calibrate the observed behaviour of the DM against the behaviour of the gas.}
			\end{centering}
\end{figure}

\subsubsection{Drag forces} \label{sec:drag}

The macroscopic behaviour of the substructure gas is determined by its macroscopic, hydrodynamic  properties. 
As the substructure's gas component moves through the cluster ICM, it experiences turbulent drag \citep{turbulence}.
The drag force on the gas
obeys the drag equation \citep{drageq},
\be
\bv{D}_\mathrm{G}=-\frac{1}{2}C\Vg A\Vg\rho^\mathrm{ICM}\Vg v^2\Vg\bv{\hat{v}}\Vg \label{eqn:Dg_expand},
\ee
where $\rho^\mathrm{ICM}\Vg$ is the density of gas in the ICM, $\bv{v}\Vg$ is the infall velocity of the substructure gas (with $\bv{\hat{v}\Vg}$ denoting the unit vector in the direction of the velocity), $A\Vg$ is its cross-sectional areas and $C\Vg$ is the coefficient of drag, which is determined by its geometry.


If the DM has a low interaction cross-section per unit mass (with respect to the in-falling gas), 
its macroscopic behaviour is instead determined by its microscopic properties (high cross-sections have been ruled out since scattering would result in evaporation and disruption of halos which hasn't been observed).
The regime of Dark Matter scattering in which we model here is a long range force, with slow momentum exchange and high preference to forward scattering, similar to that of Thompson or Rutherford scattering \citep[e.g. Mirror Dark Matter][]{mirrordm,mirrordm2}. \citet[][hereafter K13]{SIDMModel} considered such scattering, and found that the resulting interactions were frequent with a small momentum transfer in each case, resulting in an effective drag on a halo given by;
\be
\bv{D}\Vd=-\frac{1}{4}\left(\frac{\sdm}{m}\right) \rho^\mathrm{ICM}\Vd M\Vd v\Vd^2 \bv{\hat{v}}\Vd,
\label{eqn:Dd_expand}
\ee
where $\rho^\mathrm{ICM}\Vd$ is the density of DM in the ICM, $\bv{v}\Vd=\rmd\bv{r}\Vd/\rmd t$ is the velocity of the substructure DM,  $\bv{\hat{v}\Vd}$ denotes the unit vector in the direction of the velocity, $M\Vd$ is its mass, and $\sdm/m$ is the momentum transfer cross-section of the Dark Matter. 
K13 have shown that there are plausible models of SIDM which satisfy this assumption, for example interaction 
 via a dark mediator. However, they also point out that there are particle physics models of Dark Matter which could result in  evaporation of sub-halos or particle redistribution 
 and does not result in an effective drag force. We should therefore note that our model will probe specific types of anisotropic scattering due to long range forces, and not the ``hard-sphere'' SIDM with isotropic scattering that most simulators are currently modelling. In other words this observable has the potential to probe a {\em different} kind of SIDM.
 

Our self-calibrating method is based around a comparison of the forces acting on DM with those acting on the galaxies and the gas.
We therefore need to model the drag on particles anywhere between these extremes.
Equations~\eqref{eqn:Dg_expand} and \eqref{eqn:Dd_expand} provide boundary conditions: for low cross-sections, the drag force is proportional to $\sdm/m$ but, above some threshold, the force depends only on geometry of the DM substructure.
This suggests an analogue of optical 
 depth.
The coincidence that the drag is proportional to the square of velocity in both extremes is useful;
we assume that this holds throughout the transition (neglecting any phase in which the flow is laminar, and obeys Stokes' law), 
and that the DM drag force is more generally
\be
\bv{D}\Vd=-\frac{1}{2}C\Vd A\Vd\rho^\mathrm{ICM}\Vd \left(1-\mathrm{e}^{-\sdm/\sstar}\right) v^2\Vd\bv{\hat{v}}\Vd \label{eqn:map}
\ee
where geometric quantities for the DM are analogous to those for the gas, and 
\be
\frac{\sdm~}{\sstar}=\frac{1}{2\,C\Vd}\frac{\sdm/\mdm}{ A\Vd/M\Vd}. \label{eqn:sstar}
\ee
This can be interpreted as an optical depth
\be
\tau\equiv\frac{3\sdm~}{\sstar}\approx n\,\sdm s\Vd,
\ee
when $C\Vd\approx1/2$, and the substructure has a characteristic scale $s\Vd$ for which its cross-sectional area $A\Vd=\pi s\Vd^2$, $n=M\Vd/(mV\Vd)$ is the DM particle density in the substructure with a volume $V\Vd=4\pi s\Vd^3/3$,.
Figure~\ref{fig:beta} diagrammatically shows our knowledge of the two extremes between the low cross-section of the Dark Matter and the highly interacting gas and how we interpolate between the two regimes. We see that this relationship between the two regimes is essential in order to calibrate the observed behaviour of the DM to that of the gas.

Equation~\eqref{eqn:map} recovers equation~\eqref{eqn:Dd_expand} in the optically thin limit ($\tau<<1$), and is the DM analogue to equation~\eqref{eqn:Dg_expand} if it is optically thick.
The transition in behaviour occurs when the cross-section reaches a critical value $\sdm\simeq\sstar/3$
(i.e.\ $\tau\simeq1$) where, from equation~\eqref{eqn:sstar},
\ba
\frac{\sstar}{\mdm} \approx \frac{2\,C\Vd A\Vd}{M\Vd} \approx \frac{\pi s^2\Vd}{M\Vd} ~~~~~~~~~~~~~~~~~~~~~~~~~~~~~~~~~~~~~~~\\
\approx 14.1\left(\frac{s_{\mathrm D}}{100~h^{-1}\mathrm{kpc}}\right)^2\left(\frac{M_{\mathrm D}}{10^{13}h^{-1}M_\odot}\right)^{-1}\mathrm{cm}^2/\mathrm{g}~. \label{eqn:sstar_ms}
\ea


\subsubsection{Buoyancy force }
 \label{sec:buoyancy}
 
DM substructure with mean density $\rho\Vd$, moving in an ICM distributed as a singular isothermal sphere, with density $\rho_0 $ at radius $r_0$, will experience a {\it buoyancy} (MKN11);
\be 
\bv{B}\Vd=  \frac{4 GM\Vd\rho_0^2 r_0^4}{\rho\Vd s^2\Vd} \frac{\bv{r}\Vd}{r\Vd^4}.\sdm^2.
\ee
This acts in the radial direction, anti-parallel to the infall velocity, complicating our analysis.
Full hydrodynamical simulations will be essential to characterise its effect.

However, the buoyancy of DM and  gas fall off rapidly, as $\propto1/r^3$.
Such forces should be negligible outside the cluster core, furthermore, the drag according to equation \eqref{eqn:Dd_expand}  on the Dark Matter is $\propto \bv{v}^2$, and therefore will always dominate and hence, we  assume
\be
\bv{B}\Vd\approx\bv{B}\Vg\approx\bv{0}.
\label{eqn:B_expand}
\ee
In MKN11 buoyancy was assumed to be the dominant force, but here we see it can be neglected.
\begin{figure}
	\centering
	\label{fig:radial}\includegraphics[width=84mm]{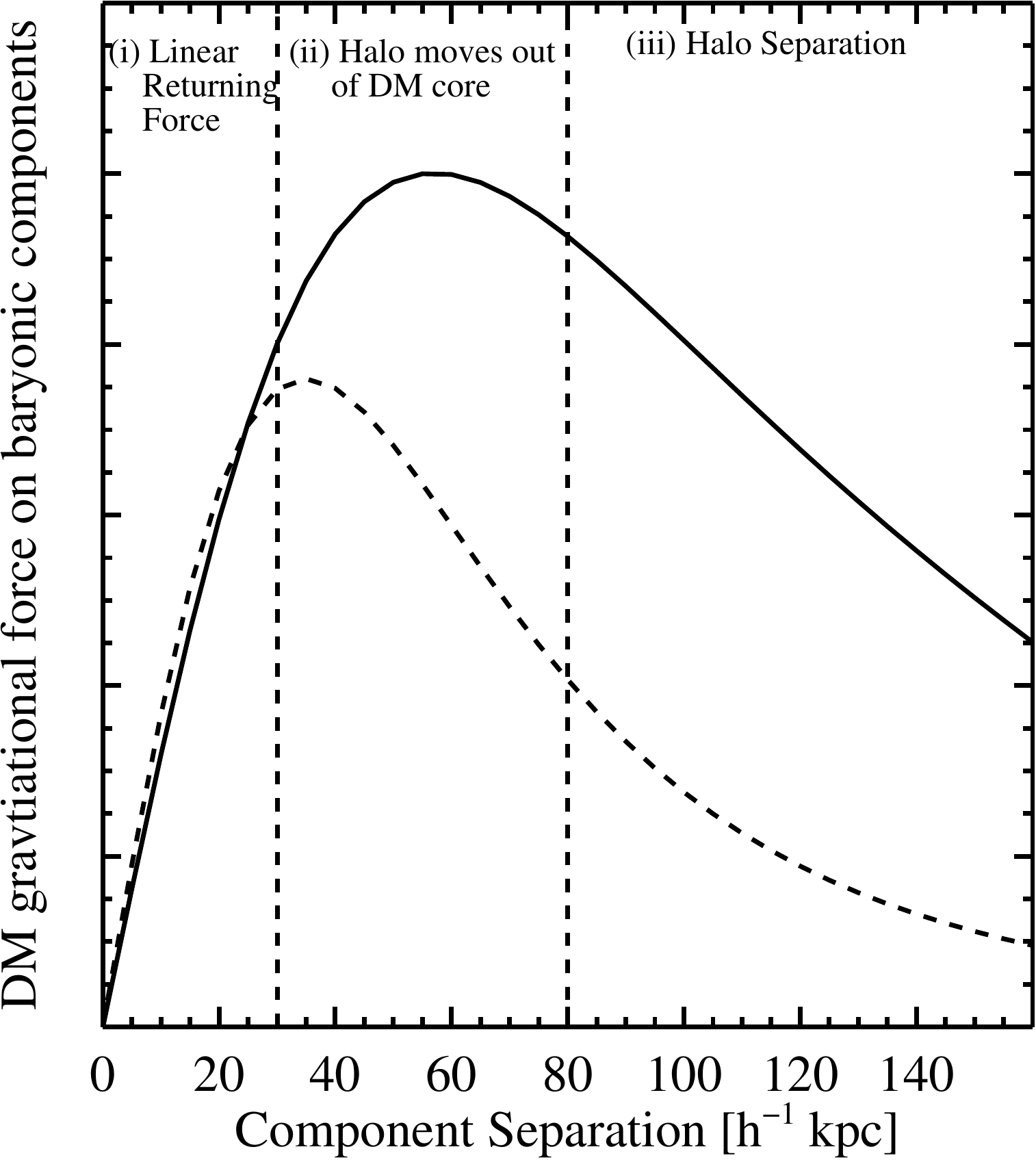}
\caption{The gravitational force that an extended Dark Matter halo has on an extended gas halo (solid line) and galaxies (dashed line), as a function of their separation.
		In this particular case, we model the force on substructure gas (solid line) and stars (dashed line) due to substructure DM, using representative component sizes discussed in the text. 
		The dashed vertical lines delineate the regimes set out in the text.
\label{fig:conv}}
\end{figure}

\subsubsection{Mutual gravitational attraction of extended substructure components}
 \label{sec:gravity}
 
The gravitational attraction of the substructure's DM acts on the gas and member galaxies to keep them bound.
WS11 commented that this force might be important if the substructure components are physically extended, but do not include it in their analysis. 
For small separations we find that it can be the most important effect.

To qualitatively understand the effect of gravitational attraction between substructure components, let us explore a simple model.
We assume the mass in each component follows a profile
\be
\rho(r)=\frac{\rho_0}{(1+(r/r_{\rm core})^2)^{3\eta/2}},
\label{eqn:profile}
\ee
where $\eta=2/3$, $r_{\rm core}=60$kpc for the DM, $r_{\rm core}=10$kpc for the galaxies and gas, and the density of the gas halo is lower by the ratio of the baryon density to the total matter density, $\rho_0^\mathrm{Gas}=(\Omega_b/\Omega_m)\rho_0^\mathrm{DM}\approx0.049\rho_0^\mathrm{DM}$ \citep{planckpars}. We model the galaxies as a delta-function.
%
and determine the force on an extended body inside a DM potential by convolving its density profile, given by equation \eqref{eqn:profile}, with the force on a point particle in the potential. 
As illustrated in Figure~\ref{fig:conv}, the force has three distinct regimes;
%
\begin{enumerate}
\item With a small separation ($\simlt 30$~kpc in this example) between two still-overlapping components, the restoring force increases linearly with separation.
\item At intermediate separations ($\sim 30$--$80$~kpc), when the components are in each others' wings, the force peaks then decreases.
\item At large separations ($\simgt 80$~kpc), the two components have separated and the force is $\propto 1/r^{2}$.
\end{enumerate}


The typically $18~h^{-1}$~kpc separations of collisionless DM and X-ray gas found by MKN11 suggest that most infalling substructure occupies the first regime, in which the substructure's three components physically overlap.
Indeed, once gas (and later perhaps DM) begins to spill out of the local potential well of the substructure's DM, they will rapidly become stripped due to tidal forces and, if they are moving fast enough near the cluster core, ram pressure.
We therefore assume that bullets {\it in which all three components are observed} must necessarily be and have always been in the first regime.

We have described above how substructure gas experiences drag from the ICM, causing it to separate from the DM; now gravity from the (dominant) DM will act to pull it back. 
The gravitational returning force increases linearly with distance from the DM in this regime, so we can model this force as
\be
\bv{G}_{\rm GD} = k_\mathrm{GD} M\Vg \bv{d}_\mathrm{GD}  = k_\mathrm{GD} M\Vg \left( \bv{d}_\mathrm{SD} - \bv{d}_\mathrm{SG} \right) ,\label{eqn:Ggd_expand}
\ee
where $k_{GD}$ is the gradient of the linear returning force.
Any drag on the DM will begin to separate it from the galaxies.
A similar gravitational restoring force will act on the galaxies, 
\be
\bv{G}_{\rm SD} = k_\mathrm{SD} M\Vs \bv{d}_\mathrm{SD}, \label{eqn:Gsd_expand}
\ee
where $k_{SD}$ is the gradient of the force opposing their separation.

We assume that the displacements of substructure components from the galaxies are antiparallel to the direction of their infall.
This is automatically satisfied if the offset is caused by the drag force. 
If buoyancy is non-negligible, or the direction of infall has changed, the offsets will temporarily display some residual component perpendicular to the direction of motion, i.e.\ finite $|\beta_\perp|$. 
Symmetry ensures that $\langle\beta_\perp\rangle=0$, but $\beta_\parallel$ may be temporarily lowered.

We have assumed that the displaced gas component will have no effect on the position of either the Dark Matter or galaxies. Indeed, in the limit that they are point particles, the gravitational attraction of the gas on its counterparts will be equal. However, since they are not the gas may act to pull the DM more than the galaxies and result in a displacement even in the case of collisionless Dark Matter. We assume that this effect is zero, but may need addressing in future experiments.

\subsection{Instantaneous quasi-equilibrium}

\begin{figure}
		\begin{centering}
			\includegraphics[width = 8cm]{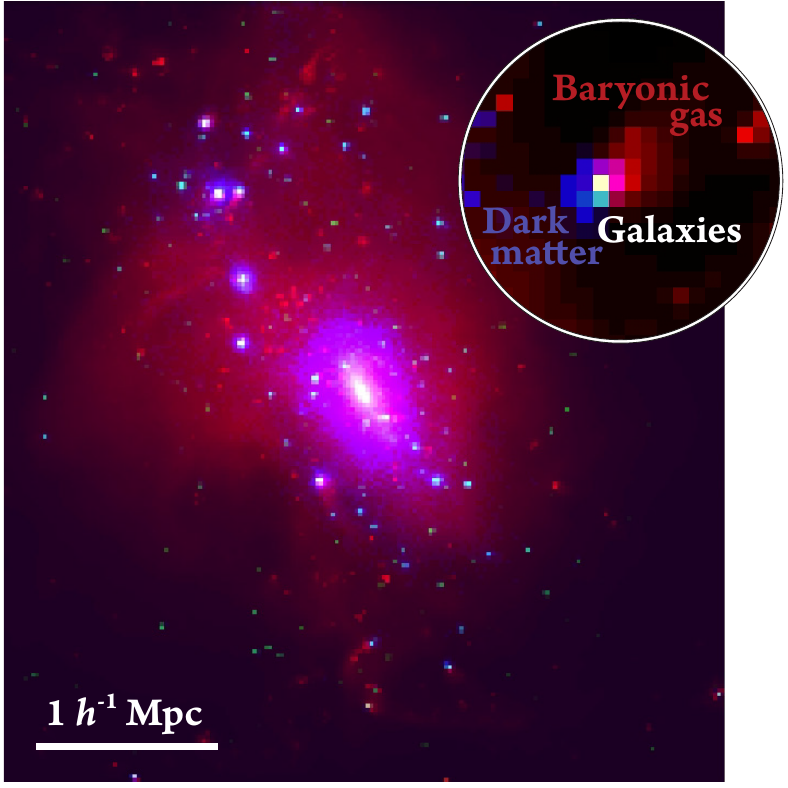}
			\caption{\label{fig:simage}
			Hydrodynamical simulation of a galaxy cluster growing through minor mergers.
			The inset zooms into one piece of infalling substructure. 
			Blue shows the projected distribution of Dark Matter, red shows the standard model baryonic gas, and white shows galaxies. 
			In this simulation, the Dark Matter is non-interacting, and therefore is expected to lie in the same place as the galaxies.
			However, there is a clear separation between infalling substructure's galaxies and baryonic gas.}
			\end{centering}
\end{figure}

We shall now consider the relative motions of the DM~(D) and gas~(G) components to the non-interacting galaxies~(S).
Any measurement of bulleticity requires observations of all three substructure components.
As discussed in Section~\ref{sec:gravity}, if the substructure ever passed very close to the cluster core, the very steep gravitational potential there would overwhelm the local substructure potential. 
Substructure gas would spill out and, unbound, would be rapidly dispersed into the ICM. 
Such a disrupted system would thus not be observed, and not enter our sample.
For substructure well away from the core
\be
\bv{r}\Vs \approx \bv{r}\Vg \approx \bv{r}\Vd \gg \bv{d}_\mathrm{SG}.
\label{eqn:large_r_approximation}
\ee
%
%
In this limit, 
and moving the galaxy frame of reference we find
\ba
\frac{\rmd^2\bv{r}\Vg}{\rmd t^2} - \frac{\rmd^2\bv{r}\Vs}{\rmd t^2} \approx
\frac{\bv{D}\Vg}{M\Vg} + \frac{\bv{B}\Vg}{M\Vg} + \frac{\bv{G}_\mathrm{GD}}{M\Vg} - \frac{\bv{G}_\mathrm{SD}}{M\Vs}
~~~~~~~~~~~~\,\\
\label{eqn:stargas_acc}
\approx 
-\frac{C\Vg A\Vg\rho^\mathrm{ICM}\Vg v^2\Vg}{2M\Vg}  \bv{\hat{v}}\Vg +
k_\mathrm{GD} d_\mathrm{SG} \bv{\hat{v}}\Vg ~~~~~ \nonumber \\
+~(k_\mathrm{SD}-k_\mathrm{GD})d_\mathrm{SD} \bv{\hat{v}}\Vd
\label{eqn:stargas_acc2}
\ea
and
\ba
\frac{\rmd^2\bv{r}\Vd}{\rmd t^2} - \frac{\rmd^2\bv{r}\Vs}{\rmd t^2} \approx
\frac{\bv{D}\Vd}{M\Vd} + \frac{\bv{B}\Vd}{M\Vd} - \frac{\bv{G}_\mathrm{SD}}{M\Vs}
\label{eqn:stardm_acc} ~~~~~~~~~~~~~~~~~~\, \\
 \approx
-\frac{C\Vd A\Vd\rho^\mathrm{ICM}\Vd v^2\Vd}{2M\Vd} \left(1-\mathrm{e}^{-\sdm/\sstar}\right)\bv{\hat{v}}\Vd \nonumber \\
 +~ k_\mathrm{SD} d_\mathrm{SD}\bv{\hat{v}}\Vd ~.~~~~~ 
 \label{eqn:stardm_acc2}
\ea
%
While the substructure passes through the outskirts of the cluster, drag separates the gas, then the DM, from the galaxies.
However, the gravitational attraction of the DM acts to pull the components back together.
The gravitational returning force increases linearly with separation, until it balances the drag and the components reach instantaneous quasi-equilibrium.
If the substructure accelerates towards the cluster, or moves through denser ICM, the drag will increase.
The components separate further, but the gravitational returning force again increases until it balances the drag force, and the system establishes a new quasi-equilibrium. 
Evidence for this equilibrium state can be seen in Figure \ref{fig:simage}, which shows an example of an in-falling sub-halo into a simulated cluster, with the red representing the gas, the blue the dark matter and the white the galaxies. We see that whilst the peaks are separated the gas has not been stripped. Without the restoring force the drag on the gas halo would cause it to separate and dissipate, and since this would be very quick the halos reach this equilibrium point. Also, the study by MKN11 found a peak separation of up to $\sim18h^{-1}$kpc, providing more evidence to show this force must balance with the restoring force. The linear scaling in Figure \ref{fig:conv} leads to the coincidence that equilibrium is
reached at the same time for the gas as for the galaxies.  That is, even if
the drag force is small which implies a slower motion toward the
equilibrium point, the distance to travel is
smaller, leading to the same time to reach equilibrium. Therefore will also reach a state of quasi-equilibrium before the halo falls in further. While in this quasi-equilibrium state, the components' accelerations and velocities are equal:
\be
\frac{\rmd^2\bv{r}\Vs}{\rmd t^2} = \frac{\rmd^2\bv{r}\Vg}{\rmd t^2} = \frac{\rmd^2\bv{r}\Vd}{\rmd t^2},\label{eqn:assumption1}
\ee
and
\be
\bv{v}\Vs=\bv{v}\Vg=\bv{v}\Vd \label{eqn:assumption2}.
\ee
Note that in this model we assume that halos retain their shape and separate. K13 found when simulating major mergers  the resulting distribution of galaxy particles post-collision is in-fact asymmetric and the peaks stay coincident. However, here we are considering smaller sub-halos in an on-going process, where particles reach a temporary equilibrium rather than a completed pass of a secondary halo in which the particles have already begun to relax. Moreover, the aim of this work is to be able interpret the weak lensing observable, which is sensitive to the mean mass distribution in a system, as a cross-section. In this sense the halos will be separated as apposed to the K13 treatment which was carried out  in the context of strong lensing which probes the peak of the mass distribution.
Under these dynamic conditions, equation~\eqref{eqn:stargas_acc2} yields
\be
d_\mathrm{SG}=\frac{C\Vg A\Vg\rho^\mathrm{ICM}\Vg v^2\Vs}{2M\Vg k_\mathrm{GD}} - \frac{(k_\mathrm{SD}-k_\mathrm{GD})}{k_\mathrm{GD}}d_\mathrm{SD} \label{eqn:dSGfin}
\ee
and equation~\eqref{eqn:stardm_acc2} reduces to
\be
d_\mathrm{SD}=\frac{C\Vd A\Vd\rho^\mathrm{ICM}\Vd v^2\Vs}{2M\Vd k_\mathrm{SD}} \left(1-\mathrm{e}^{-\sigma/\sstar}\right). \label{eqn:dSDfin}
\ee
We shall assume that the baryon fraction 
\be
f_b\equiv\frac{\Omega_b}{\Omega_m}=\frac{\rho^\mathrm{ICM}\Vg}{\rho^\mathrm{ICM}\Vd+\rho^\mathrm{ICM}\Vg}
\ee
is roughly constant throughout the system. This implies
\be
\frac{\rho^\mathrm{ICM}\Vd}{\rho^\mathrm{ICM}\Vg}\approx\frac{M\Vd}{M\Vg}=\frac{1-f_b}{f_b}.
\ee
This approximation may not be quite accurate if the baryon faction depends on the radius from the cluster centre, however, this is a conservative estimate and should not result in an over-estimate of $\sdm/m$.
We shall also assume geometric similarity so the drag coefficients coincide
\be
C\Vd\approx C\Vg ,
\ee
as do the areas
\be
A\Vd\approx A\Vg ,
\ee
and
\be
k_\mathrm{GD}\approx k_\mathrm{SD}.
\ee
The former is a conservative estimate since we predict that the shape of gas halos will be more streamline than DM halos (due to tidal stripping), resulting in a larger drag for on the halo.The latter is reasonable because the values of $k$ are mainly driven by the inner slope of the same DM potential.
However, the values are also perturbed by the distribution of mass in the gas and galaxies, so it may be necessary to model for future surveys, when averaging over many thousands of clusters allow a high-precision measurement.

Taking the ratio of equations~\eqref{eqn:dSGfin} and \eqref{eqn:dSDfin}, we find
\be
\beta\equiv\frac{d_\mathrm{SD}}{d_\mathrm{SG}}\approx
1-\mathrm{e}^{-\sdm/\sstar} \label{eqn:beta_complicated} .
\ee
Hence we find that our proposed quantity is independent of the substructure infall velocity and the time since the infall began. 

Recall from equation~\eqref{eqn:sstar_ms} that $\sstar$ strictly depends upon the size and mass of each piece of substructure. 
When we come to compute and interpret a mean value $\langle\beta\rangle$, it might be necessary to measure these properties and weight measurements from each piece of substructure appropriately, or to constrain and statistically marginalise over a distribution of $s\Vd^2/M\Vd$ with global nuisance parameters.
This may be necessary for future, high-precision measurements using many thousands of clusters.
To interpret the first observations of this effect, it should be sufficiently accurate to assume a mean value $\langle s\Vd^2/M\Vd\rangle\sim4.5\,\mathrm{cm}^2/\mathrm{g}$.


\begin{figure*}
	\begin{minipage}{180mm}
		\begin{centering}
			{\label{fig:SGr}
			\includegraphics[width = 5.5cm]{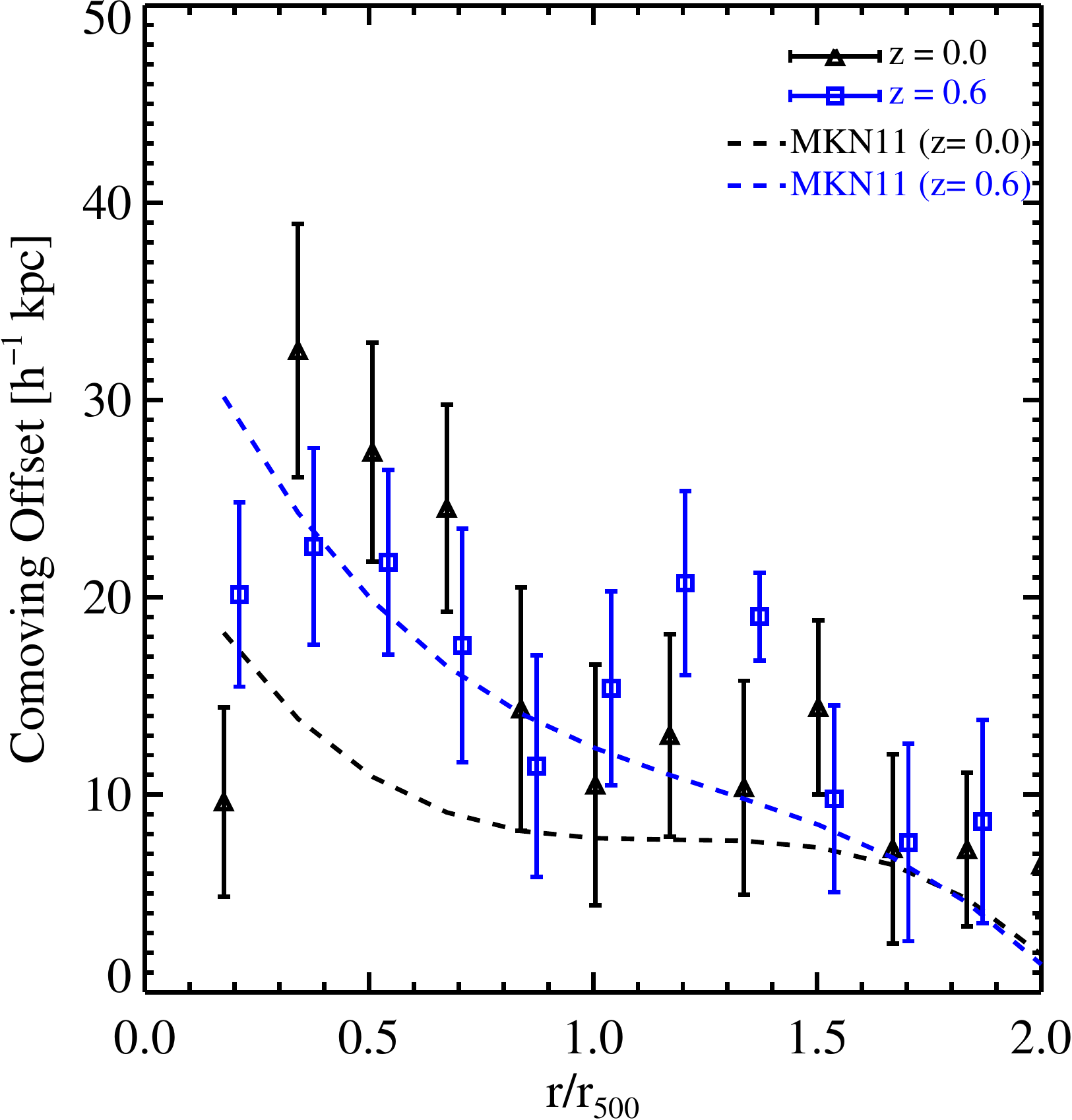}}
			\qquad
			{\label{fig:SGm}
			\includegraphics[width = 5.5cm]
			{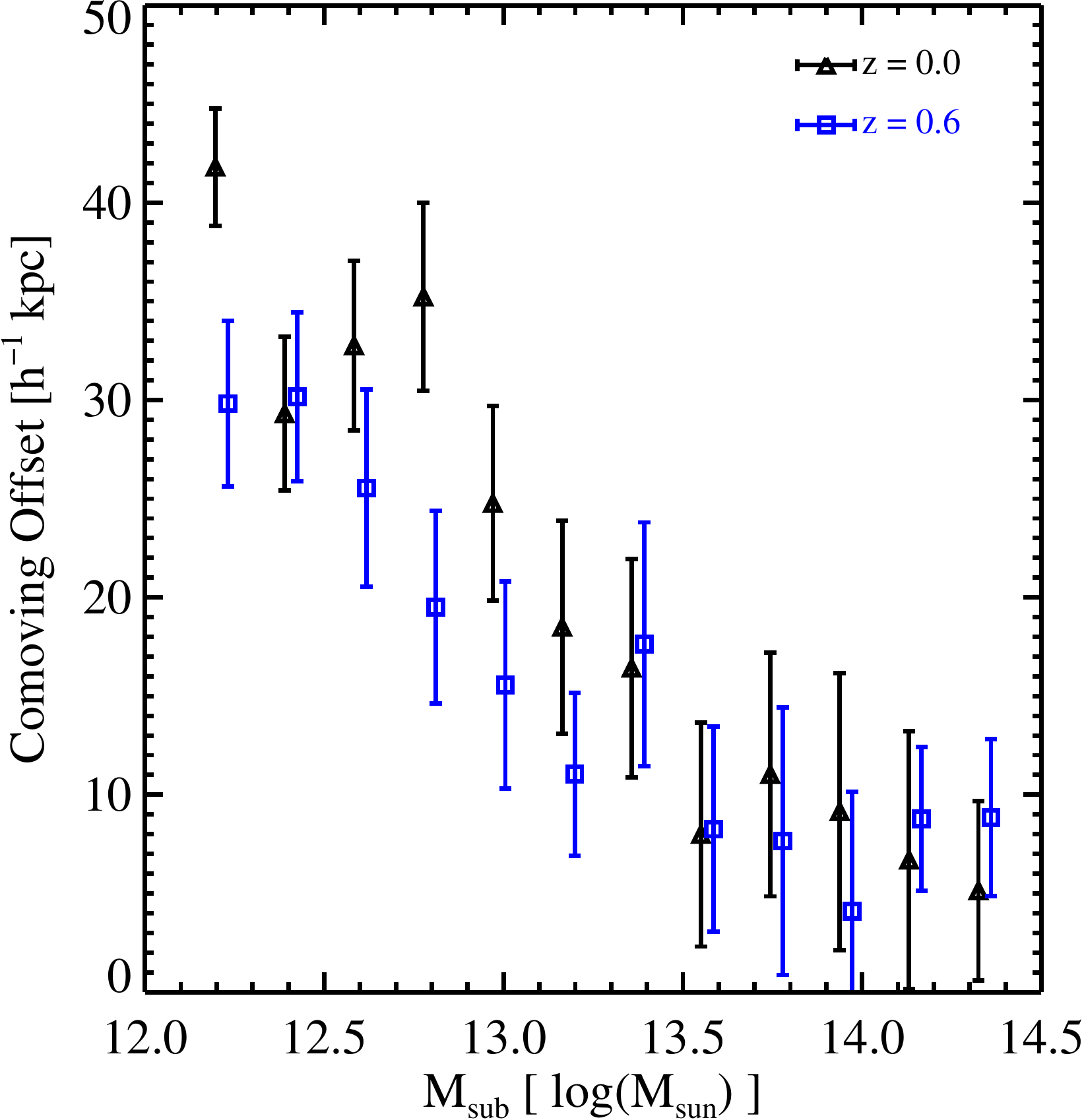}}
			\qquad
			{\label{fig:SGp}
			\includegraphics[width = 5.5cm]{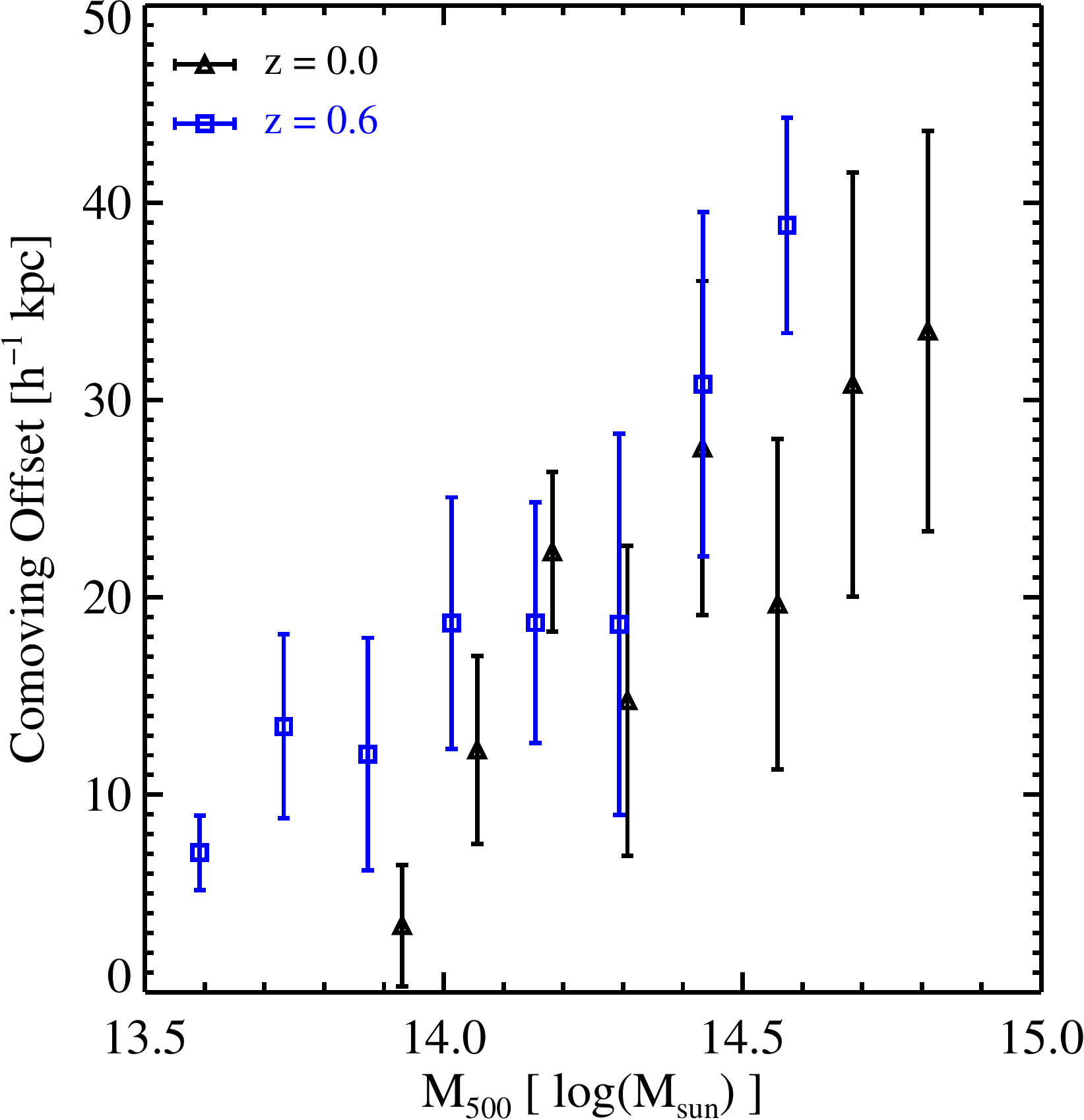}}
			\caption{\label{fig:SG}
			Projected offsets $d_\mathrm{SG}$ between galaxies and gas in substructure around 30 clusters in hydrodynamical simulations that include the effects of gas cooling, star formation, supernova feedback and AGN feedback. In each case the black points are the results from clusters at a redshift of $0$ and the blue points are from halos at a redshift of $0.6$.  The left panel shows the offset as a function of the projected distance from the cluster in units of $r_{500}$, the radius inside which the density is 500 times greater than the mean density in the Universe. The dashed lines show earlier predictions from MKN11. The centre panel shows the offset as a function of the mass of the sub halo (see equation \ref{eqn:msub}). The right panel shows the offset as a function of its parent cluster mass, $M_{\rm 500}$, the total mass inside a sphere of radius $r_{500}$. Each point shows the weighted mean of offsets within that particular radial or mass bin, with the error bars representing the one-sigma error.}
			\end{centering}
		\end{minipage}	
\end{figure*}

\section{Applying the method to simulations}\label{sec:results}

\subsection{Hydrodynamical simulations of clusters}\label{sec:sims}

To check the feasibility of measuring $\sdm/m$ in real astronomical data, 
we need to apply our method in a controlled environment,
hydrodynamical simulations of galaxy clusters.
 These simulations use non-interacting DM with 
$\sdm=0$, so they will be useful only to predict the typical level of 
signal-to-noise ratio for observations.

We study 30 galaxy clusters, extracted from a large ($[500 h^{-1}{\rm Mpc}]^3$) Dark Matter only simulation, run as part of the Virgo Consortium's 
Millennium Gas project (Pike et al. in preparation). These were re-simulated 
using Gadget 2 \citep{gadget2}, where the gas dynamics is modelled using the 
Smoothed Particle Hydrodynamics (SPH) method. The WMAP 7 cosmology was 
adopted \citep{WMAP7} with $\Omega_{\rm m}=0.272$, $\Omega_\Lambda=0.728$, 
$\Omega_{\rm b}=0.0455$, $h=0.704$ and $\sigma_8=0.81$. Clusters were selected
by defining 5 bins equidistant in $\log(M_{200})$, between 
$10^{14} h^{-1}M_{\odot}$ and $10^{15} h{^-1} M_{\odot}$, and drawing 6
objects at random from within each bin. The mass resolution was chosen to
keep the number of particles constant ($\sim 10^6$) within $r_{200}$, such that the dynamic 
range of cluster substructure was similar across the mass range.  
Furthermore the spatial resolution ranged between $3-8 h^{-1}$ comoving kpc and $9-15 h^{-1}$ comoving kpc for redshifts z=0.6 and z=0 respectively.
The gravitational
softening length (held fixed in physical co-ordinates at $z<3$) was set to 
$\epsilon=6 h^{-1}{\rm kpc}$ for the most massive haloes, decreasing to $3 h^{-1}{\rm kpc}$
for the least massive objects.

Radiative cooling (assuming zero
metallicity gas), star formation and feedback from stars and active galactic
nuclei (AGN) were implemented, as described in \citet{starform}. Including AGN
feedback is particularly important for avoiding a cooling catastrophe and
broadly reproducing the observed cluster scaling relations at low redshift.
The prescription used for the simulations follows that set out by \citep{AGNfeedback}. Black hole seeds were inserted at a redshift of 5.2, where a 
gas particle was converted in each subhalo or friends-of-friends (FOF) group 
with $M_{200}>3\times10^{10} h^{-1}M_\odot$, where $M_{200}$ is the mass within the radius at which the mean density is 200 times greater than the mean density in the Universe. 
Each black hole had an initial mass of $10^5 h^{-1} M_{\odot}$ and could subsequently grow via mergers with
other black holes or accretion of gas using a modified version of the 
Bondi Hoyle formula. The available energy for feedback was proportional to the
mass accreted onto the black hole, with an overall heating efficiency of
1.5 per cent. Gas particles were heated to a fixed temperature (varying from
$10^8$K in the lowest mass clusters to $10^{8.5}$K in the most massive systems)
when the required amount of energy was available. 

For each of the 30 clusters, we constructed projected 2D maps of the density of 
the Dark Matter, the stellar material and the hot ($T>10^{6}$K) X-ray emitting 
gas along the z-axis. For the analysis we observed the clusters at 
two snapshots; one at a redshift $z=0.6$ and the other at redshift zero. 
At these redshifts, the 30 clusters have $M_{500}$ masses spanning the range 
$10^{13.5}$ -- $10^{14.7}M_\odot$, with a mean mass 
$2.6\times10^{14} M_\odot$ and $1.1\times10^{14} M_\odot$ at $z=0$ and 
$z=0.6$ respectively, where $M_{500}$ is the mass within the radius at which the mean density is 500 times greater than the mean density in the Universe. Fig.~\ref{fig:simage} shows the density field 
from one of the simulated clusters at $z=0.6$. Here, the distribution of DM is
shown in blue, the hot gas in red and the stellar material (galaxies and 
intracluster light) in white. The inset shows a zoomed view of a typical 
piece of subtructure where the DM and gas are clearly separated. 

We use the public code {\sc Wavdetect}, from {\sc Ciao tools} 
\citep{CIAOWavdetect} to identify peaks in the DM, gas and stellar density 
maps. With our better peak detection algorithm than MKN11, and a better model 
for AGN physics, we are now able to include substructure anywhere near a 
cluster, including the inner core ($r<0.3~r_{500}$). The substructure masses 
span the range $10^{12.0}$-- $10^{14.4}{\rm M}_\odot$ with a mean mass 
$8.6\times10^{12} {\rm M}_\odot$ and $6.3\times10^{12} {\rm M}_\odot$ at 
redshifts $z=0$ and $z=0.6$. We find, on average, $10$ substructures per 
cluster, with a mean value of 
$\langle M_\mathrm{sub}/M_\mathrm{cl}\rangle=0.03$.
We match adjacent gas, DM and stellar mass peaks, recording the positions and 
the standard errors returned by {\sc Wavdetect}. For now, we complete this 
process without noise, but we shall repeat it in the presence of realistic 
observational noise in Section~\ref{sec:data}.
comoving coordinates.

\subsection{Component offsets in noise-free simulations}

The mean offset between substructures' galaxies and baryonic gas is shown in Figure~\ref{fig:SG}, as a function of various cluster properties.
Position estimates from low-mass peaks are noisy, so we use inverse variance error estimates to compute a weighted mean.
The black (blue) points show the offset around clusters at redshift $z=0$ ($z=0.6$).

The left panel of Figure~\ref{fig:SG} shows the offset between substructure components (in units of $h^{-1}$kpc), as a function of projected distance from the cluster in units of $r_{\rm 500}$. 
The dashed lines show the results of MKN11 as reference.
At a redshift $0.6$, we recover a similar $\sim20h^{-1}$kpc offset, but we find no statistically significant redshift dependence. 
We find that the offset drops at small projected radii. 
This is probably because substructures really passing through the core are disrupted and dispersed. 
We therefore preferentially see substructure at large 3D radii, whose positions have been projected near the centre of the cluster.
Their separations align nearly with the line of sight, so their projected separations appear small. 

The middle panel of Figure~\ref{fig:SG} shows the offset as a function of the substructure mass.
To estimate the substructure mass, we used the ratio of the total mass 
signal detected by {\sc Wavdetect} near the substructure ($S_{\rm sub}$) and main cluster ($S_{\rm cl} $), i.e.
\be
M_{\rm sub} = \frac{S_{\rm sub} }{S_{\rm cl} }M_{\rm cl},
\label{eqn:msub}
\ee
where $M_{\rm cluster}$ is the mass of the main cluster $M_{\rm 500}$. 
The decreased offset for massive substructure is consistent with our analytical model.
The larger gravitational returning force will bind the stars and gas closer to the DM throughout infall.

The right panel of Figure~\ref{fig:SG} shows the offset as a function of the parent cluster mass. 
The increased offset near massive clusters is also consistent with our analytic model. 
More massive clusters have a higher density ICM, so the drag (and buoyancy) forces that drive the offsets will be increased.

\begin{figure}
		\begin{centering}
			\includegraphics[width = 7.5cm]{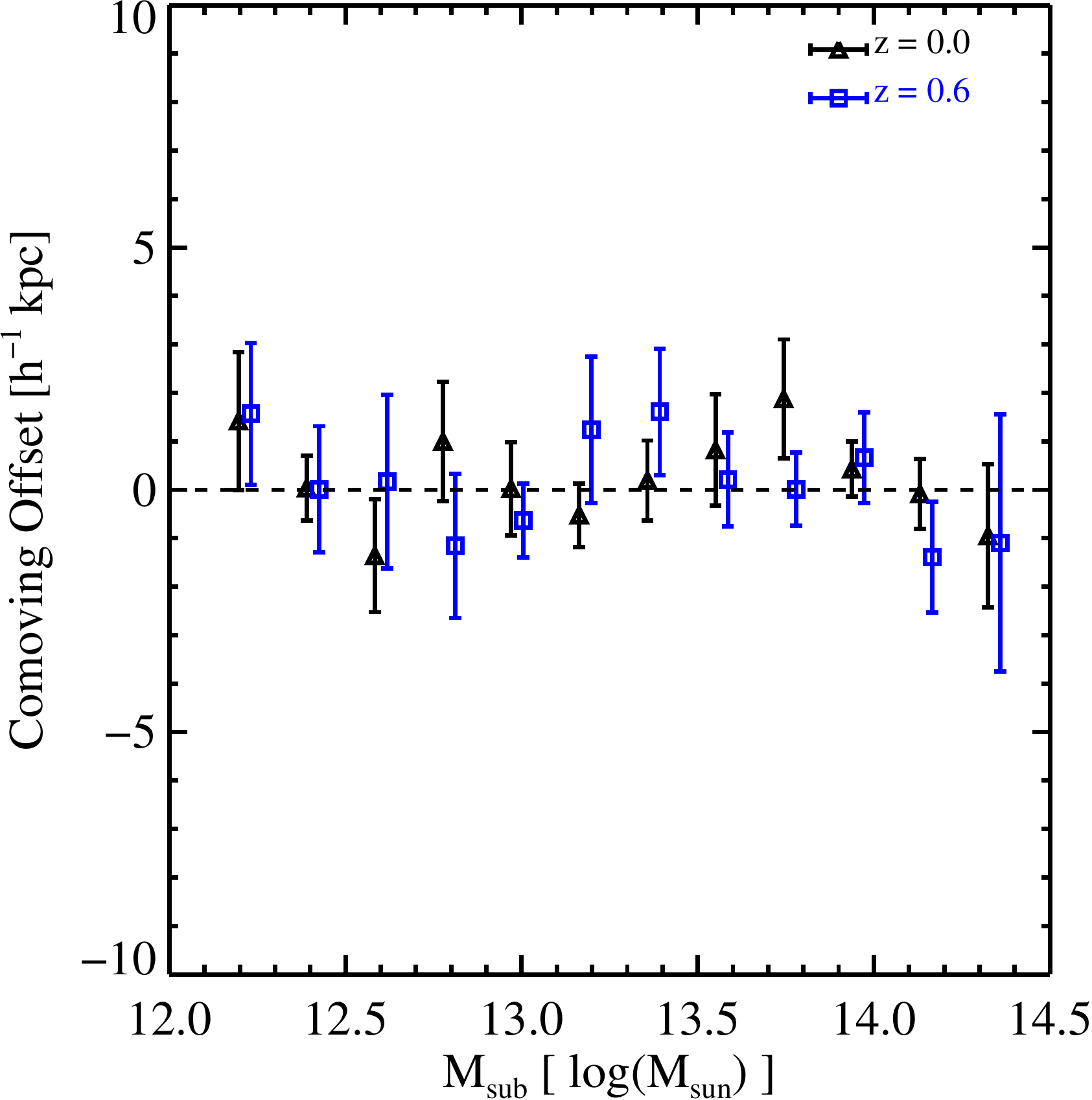}
			\caption{\label{fig:SI}
			Projected offsets between substructure galaxies and the intersection point with DM in the direction towards the gas (SI) for $z=0$ and $z=0.6$. In real data, this distance will probe the finite cross-section of DM. Since our simulations explicitly use collisionless DM, we expect these offsets to be consistent with zero. Each point shows the weighted mean of offsets within that particular mass bin, with the error bars representing the one sigma error.}
			\end{centering}
\end{figure}

We next look at the offset between substructures' galaxies and DM.
In real data we expect this offset to reflect the interaction cross-section of DM, and any detected offset will imply a non-zero $\sdm/m$.
The DM used for these simulations is  collisionless, so we expect the offset to be consistent with zero.
The offset between galaxies and the DM intersection point is shown in Figure~\ref{fig:SI}.
The position of the DM is indeed consistent with that of the member galaxies at both redshifts.
Since there is also no significant gradient towards low mass substructure, we are confident that there is no residual bias in the simulations or subsequent analysis. 

\begin{figure}
		\begin{centering}
			\includegraphics[width = 7.5cm]{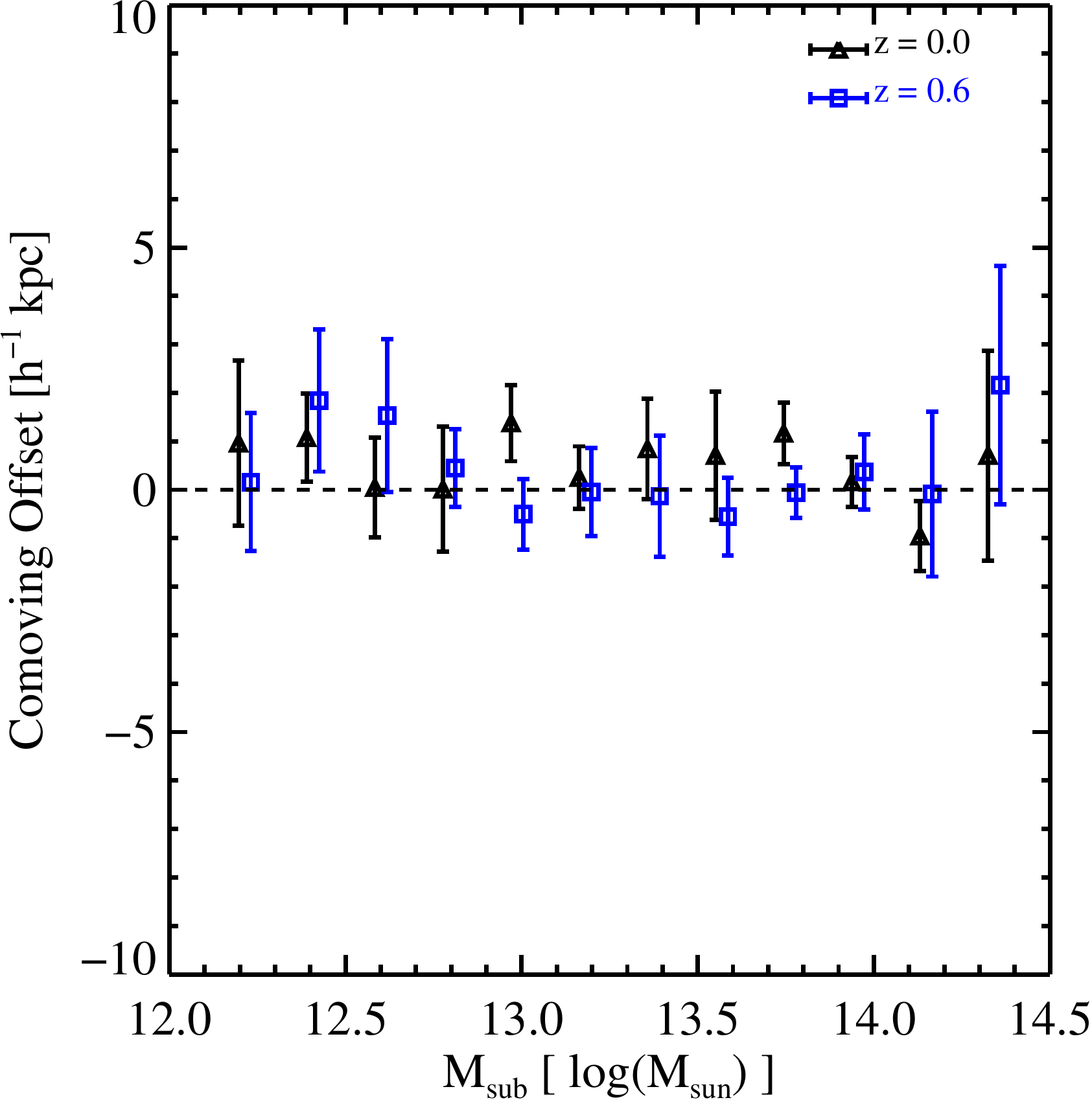}
			\caption{\label{fig:DI}
			Projected, transverse offsets between DM and the intersection point (DI) for $z=0$ and $z=0.6$, which tests for potential systematics. Under the assumption that over an ensemble average there is no preferred in-fall direction and there is no systematic bias in the positional estimates of DM, this parameter should be consistent with zero. Each point shows the weighted mean of offsets within that particular mass bin, with the error bars representing the one-sigma error.}
			\end{centering}
\end{figure}

The transverse distance between the DM position and the intersection point, $d_{\rm DI}$, reflects the error in the estimated position of the DM, and should therefore be unbiased and consistent with zero. 
Figure \ref{fig:DI} shows the offset between the DM and the intersection point as a function of the mass of the sub-halo.
We see that the offset is consistent with zero in all cases, even at lower signal peaks. We are therefore confident there is no residual bias in the simulations or analysis.

\begin{figure*}
	\begin{minipage}{178mm}
		\begin{centering}
			{\label{fig:GI10comp}\includegraphics[width = 7.5cm]{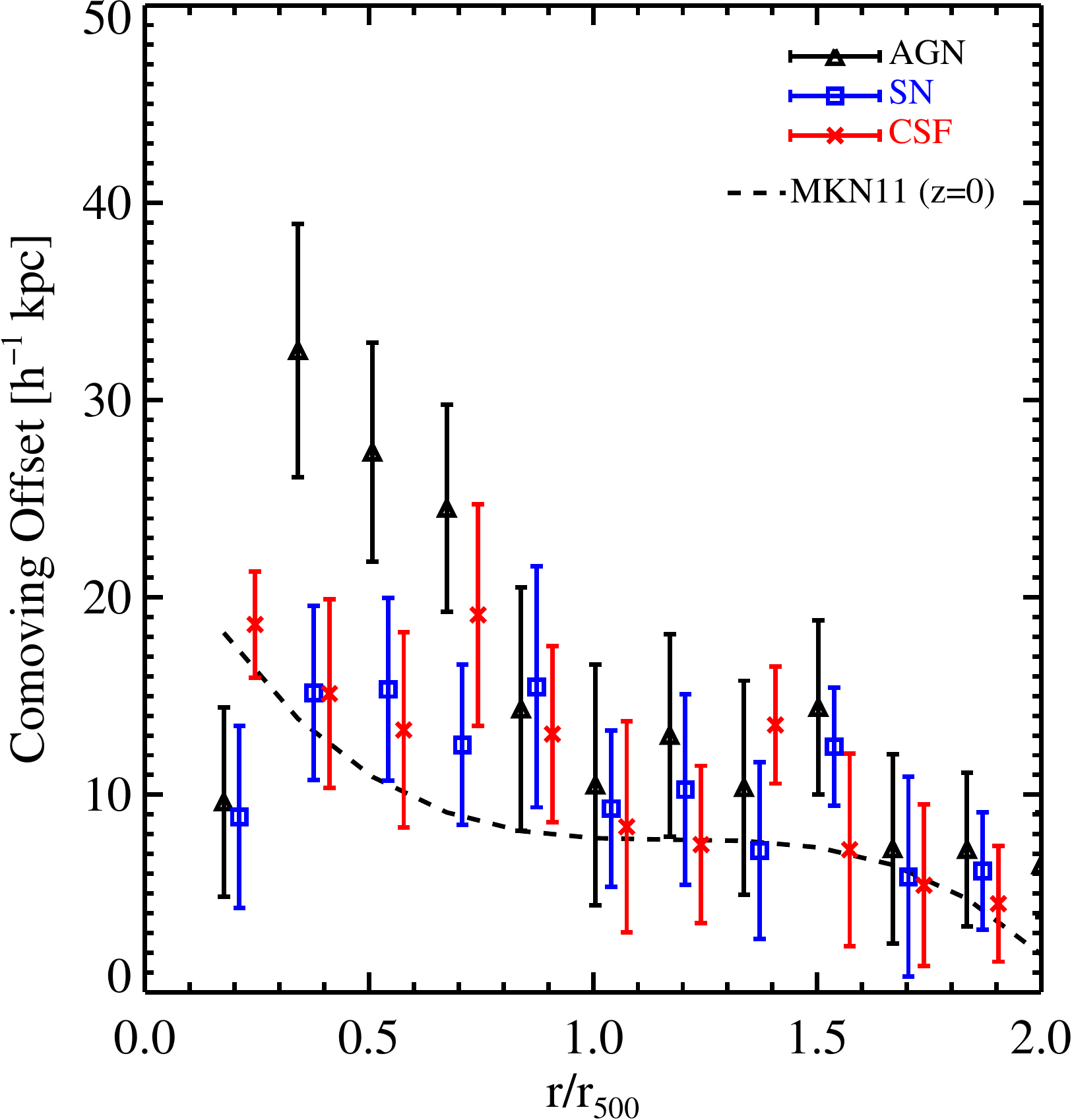}}
			{\label{fig:GI06comp}\includegraphics[width = 7.5cm]{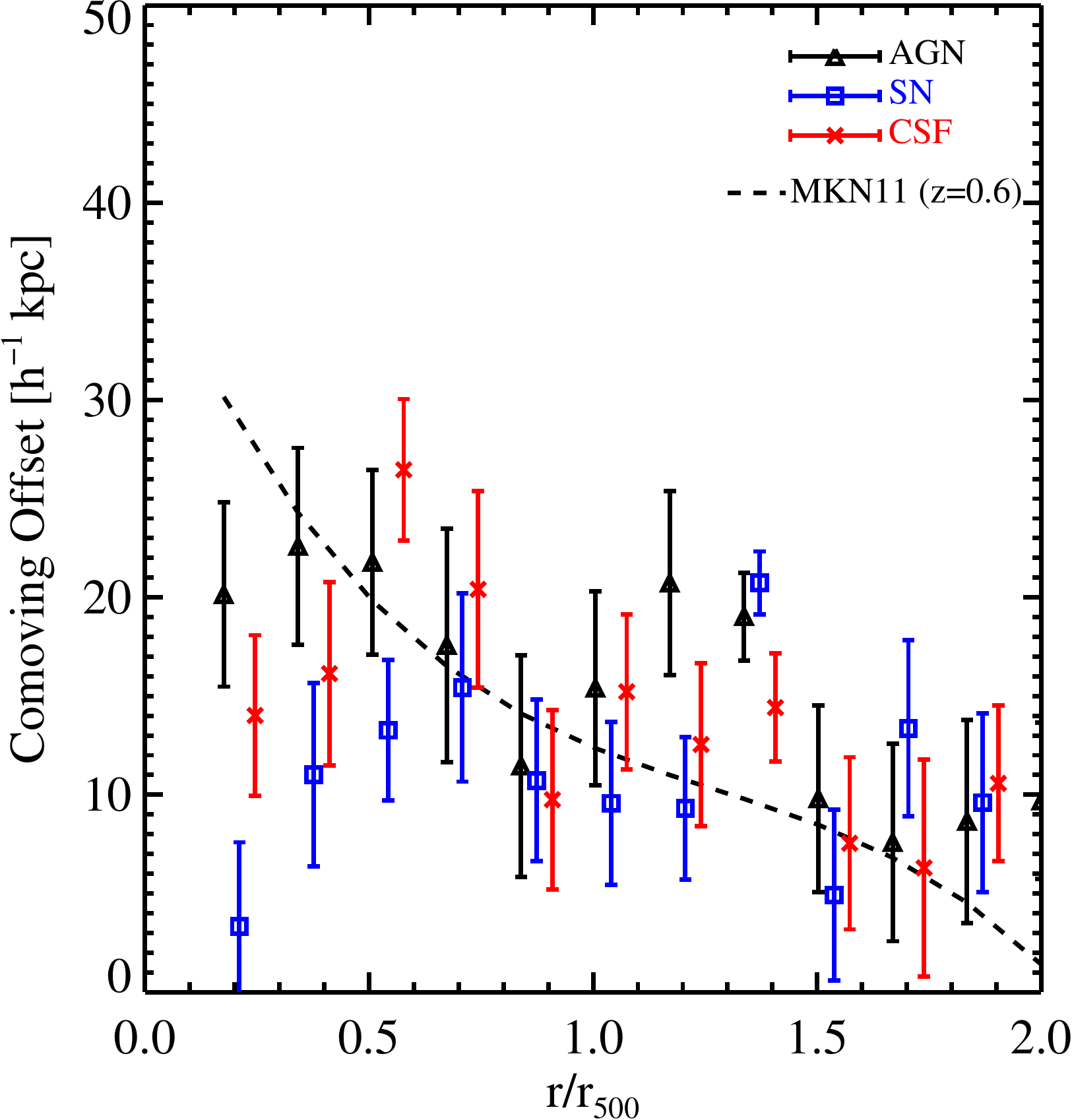}} \\
			~~~~~\qquad~~~~~
			\caption{\label{fig:comp}
			The observed offsets in substructure components appear broadly robust to astrophysical processes.
			The panels show the projected offsets $d_\mathrm{GI}$ between gas and Dark Matter (in the direction towards the galaxies) in substructure around 30 simulated clusters at $z=0$ (left) and $z=0.6$ (right), assuming different models of baryonic physics. 
			Dashed lines show earlier predictions from MKN11, for reference and ease of comparison.}
			\end{centering}
		\end{minipage}	
\end{figure*}
\subsection{Robustness to astrophysical effects}

We find the offsets between substructure components at $z=0$, consistent with those reported by MKN11.
However, we find little evolution with redshift, which was found by MKN11, with discrepant offsets by $z=0.6$.
The different baryonic physics included in the two simulation codes may account for this. 
As in this paper, MKN11 modelled the cooling of gas, star formation and supernova feedback but we also include feedback from AGN. 
AGN have a prominent effect throughout the cluster environment that may change the dynamics and properties of the in-falling sub-halos. 

To test the effects of astrophysical processes on substructure offsets, we repeat our analysis on a range of simulations.
Figure \ref{fig:comp} shows the offset between substructure gas and DM at $z=0.6$ (left) and $z=0$ (right) in simulations with varying degrees of baryonic physics; 
\begin{enumerate}
\item Cooling and star formation only (CSF, red points in Figure \ref{fig:comp}).
\item Cooling, star formation and supernova feedback (SN, blue points in Figure \ref{fig:comp}).
\item Cooling, star formation, supernova and active galactic nuclei feedback (AGN, black points in Figure \ref{fig:comp}).
\end{enumerate}

The offset signal remains measurable in all cases, but we find differences near the cluster core and especially at low redshift.
This is presumably due to the injection of outward energy by AGN into even the substructure gas,
and will be more evident at lower redshift since feedback is proportional to the square the black holes mass which is increasing with cosmic time.
The simulations without AGN feedback (blue points) are more consistent with the MKN11 simulations of similar physics. 
We therefore conclude that the small discrepancy between the amplitude of the offsets reported in this paper and MKN11 are potentially due to different prescriptions of baryonic physics.
Indeed, the discrepancies could be attributed to how different codes simulate baryonic physics. In MKN11, the code used to simulate the bayonic gas physics in the cluster was an adaptive mesh refinement code (AMR), whereas the code used for the simulations in this work was based on smooth particle hydrodynamics (SPH). 
It is commonly understood that AMR and SPH differ in the way they calculate the physics of hydrodynamical bodies in clusters. These differences are most evident in the apparent stability of gaseous halos. 
AMR structures are more likely to disrupt and disperse that SPH gas halos \citet{AMRvSPH}. This could mean that any in-falling gas halo that separates from its bound DM sub-halo may disperse before exhibits a large separation from its DM host. On the other hand, SPH is know to form much more stable structures, which may mean that halos can become significantly more separated before it disrupts and becomes part of the ICM. These differences could contribute to the discrepancies in displacements between the work in MKN11 and this study. We note  that although there are differences between the two studies, the true underlying nature of the simulations and resulting implications for cluster dynamics are beyond the scope of this paper.

\subsection{Total matter vs.\ Dark Matter systematic bias?}{\label{sec:bias}

We shall advocate using gravitational lensing to map the distribution of DM.
However, gravitational lensing probes the {\it total mass} along a line of sight (see reviews by \citealt{BS01, RefregierRev, MKRev}).
The total mass is dominated by DM, but roughly $15\%$ is in the baryonic gas at an offset location.
In their analysis of the Bullet Cluster, \citet{separation} fitted the distribution of mass due to the X-ray emitting gas, and subtracted that from the lensing measurement of total mass before . 
In principle, it would be possible to do the same in minor mergers, although the much lower S/N may cause practical difficulties.

In order to test such a systematic we convert the projected density fields of the cosmological simulations used in section \ref{sec:results} into 30 gravitational lensing maps via the formalism in \citet{KS93}, which demonstrates how the lensing signal is related to the projected surface density via a convolution. We limit our mock observations to the field of view of the \textit{Hubble Space Telescope} (HST) Advanced Camera for Surveys, and assume a density of $80$ galaxies ${\rm arcmin}^{-2}$, as expected from a two-orbit exposure using the F814W band. We then scatter the background galaxies randomly on the sky, and interpolate the shear field to their positions, assume they are all $z_\mathrm{source}=1$. From these shear maps we conservatively select the two most massive DM halos and reconstruct the expected position (with \textit{no noise in the ellipticity of the galaxy}), for two cases; one with the total matter in the simulation and one with the Dark Matter only.  Then using 
 {\tt LENSTOOL} \citep{lenstool}, we reconstruct the DM positions with flat priors centred on the distribution of galaxies, making sure this prior includes the position of the gas halo. We test the effect of including the full matter distribution by calculating the resultant $\beta_\parallel$ using equation \eqref{eqn:alphaparallel} for the two cases. 

As Figure~\ref{fig:dist_PDF} shows, the false assumption that lensing measures only DM does indeed introduce a bias, mimicking the effect of a small interaction cross-section -- but at the very low level of $\Delta\beta_\parallel\sim0.005$.
This is an order of magnitude below the statistical accuracy that will be possible with existing data (see next Section), so we shall neglect the effect for now.
We adopt the algorithm for Dark Matter astrometry by \citet{Harvey13}, which deals with all other potential sources of bias.
When the method is applied to future, very large surveys, and probes $\beta_\parallel<0.05$, we suggest that the algorithm should be extended to simultaneously fit the DM and gas mass.

\section{Prospects for measuring signal with realistic noise}\label{sec:data}

By using the location of substructure galaxies as a
proxy for the direction of in-fall, one can retain the $\sim20h^{-1}$kpc absolute offset between DM and gas seen by MKN11.
To estimate this {\it signal} as accurately as possible, we have so far exploited noise-free simulations, and used many substructures per cluster \cite[using the well-known $\Lambda$CDM tendency to produce more satellite halos than observed in data, e.g.][]{CDMsatellites}.
To estimate a realistic {\it signal-to-noise}, and the prospects for constraining $\sdm/m$, we shall now add observational noise reflecting existing datasets to the shear fields used in section \ref{sec:bias}.

Once again, identifying only the two most massive sub-halos in each of our sample of 30 clusters, we then consider the expected noise on the positions of each of their components.


\begin{figure}
		\begin{centering}
	 		\includegraphics[width = 8cm]{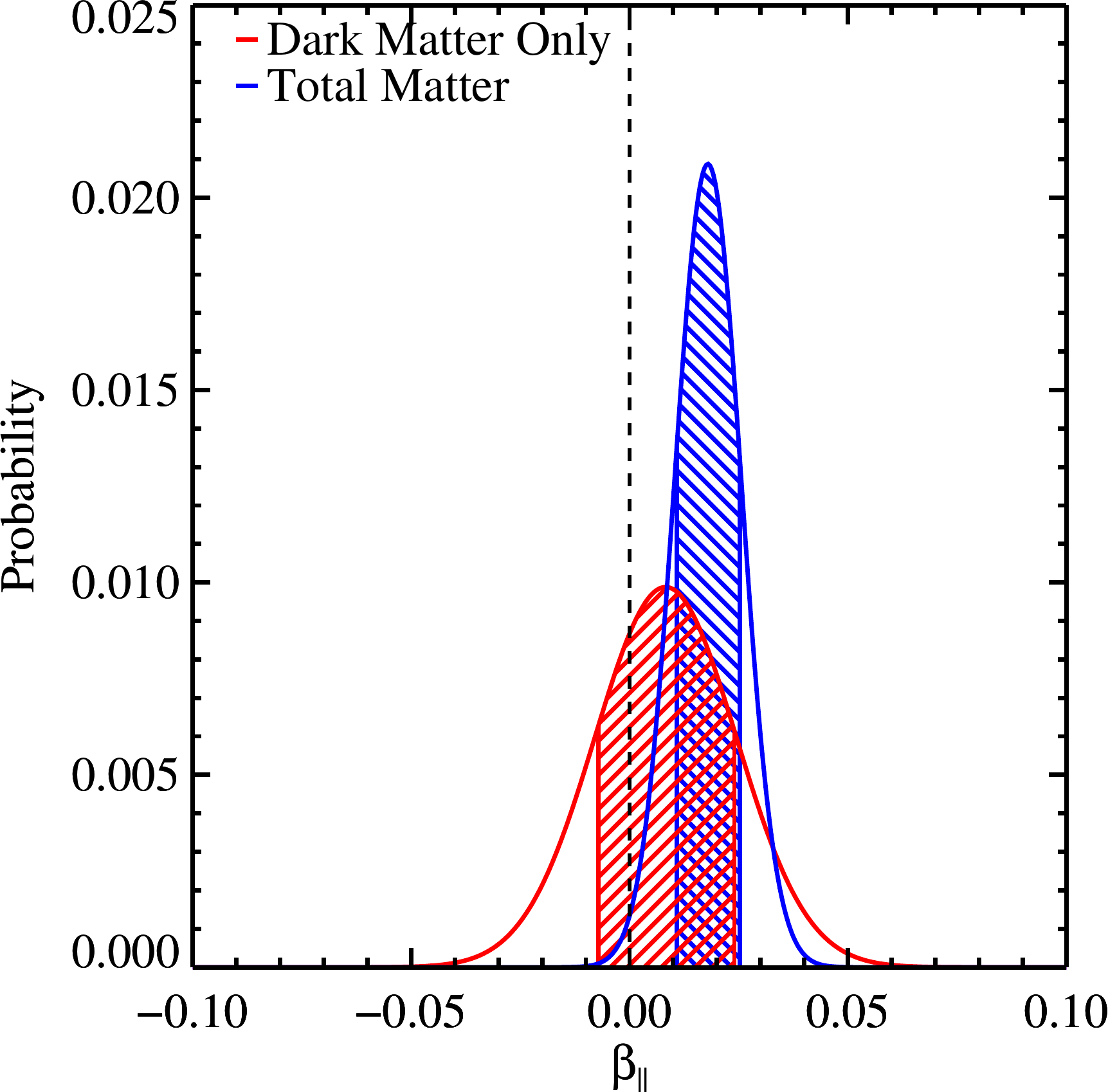}
			\caption{\label{fig:dist_PDF}
Bias induced in measurements of $\beta_\parallel$ by assuming that gravitational lensing measures only DM (blue), rather that the total mass (red).
This is {\it an order of magnitude below} the expected statistical precision for 30 clusters.
We can only detect this systematic effect in our {\it noise-free} simulations -- note the change of scale on the horizontal-axis is compared to Figure~\ref{fig:dist_redshift}.
Only in very large, future surveys, will it be necessary to simultaneously fit (and subtract) the mass in the other substructure components.}
			\end{centering}
\end{figure}

\subsection{Signal as a function of cluster redshift}
 
Our simulations show no redshift dependence in the various substructure component offsets,
in units of physical comoving separation (Figure \ref{fig:SG}).
However, the apparent {\it angular} offsets, and the amount of a cluster visible in a telescope's field of view, will depend upon the distance to the cluster
(its gravitational lensing signal also depends on the distance).
We could rely upon this lack of evolution.
However, to compare measurements from clusters at different redshifts in a controlled way, we
take the mass snapshot of each cluster at $z=0$, and rescale it as if it were at $z_{\rm lens}=0.2$, $0.4$ and $0.6$.
In each case, we assume all 30 clusters are at the same redshift; in reality, a sample will include clusters from a range of redshifts.


We  
impose an source galaxy intrinsic ellipticity distribution with $\sigma_\epsilon=0.3$ 
\citep{COSMOSintdisp}.
Previous work showed that other sources of error in a lensing analysis are subdominant \citep{Harvey13};
here we are primarily 
interested in the potential for error from 
the complex morphologies of our simulated substructure.
For those halos that exist at a redshift of z=0.6, however lie outside of the field of view (FOV) at a redshift of z=0, we discard these and select the next largest sub-halo in the FOV.

\subsection{Noise in X-ray and Galaxy observations}

We note that the noise in the positions of the member galaxies should be sub-dominant to the error in the position of the Dark Matter. 
The error in the gas halos however, should be reasonably well-estimated from the simulations. 
In the case of isolated sub-halos in observational X-ray data, positions can be extremely well constrained. The dominant error in the positions of gas halos will arise from estimating the distribution of non spherical, amorphous halos that have uncertain merger histories. 
Such effects are simulated via the hydrodynamics of the simulation and are included in the positional estimates from our peak finder however to reflect expected shot noise we introduce additional noise into the X-ray halo positions.

\subsection{Expected statistical precision}

After measuring the position of all the substructure components, we measure $\beta_\parallel$ (see equation~\ref{eqn:alphaparallel}) for each bullet.
We model the probability distribution function (PDF) of this as a Gaussian centered on the best-fit value and a width corresponding to the measurement error.
Figure \ref{fig:dist_redshift} shows the stacked PDF of $\langle\beta_\parallel\rangle$ at three different redshifts. 

All of our estimates of $\beta_\parallel$ are consistent with zero, as expected for simulations of non-interacting DM (Figure \ref{fig:dist_redshift}).
Constraints are tightest for clusters at low redshift, where the angular separation of components is larger, and the gravitational lensing signal is stronger.
For clusters at $z=0.2$, the two-tailed 68\% confidence limit on $\langle\beta_\parallel\rangle$ is $\pm0.15$,
inferring that we will be able to make a $\sim6\sigma$ detection of an offset between DM and baryonic gas in data.

To estimate the constraints on $\sdm/m$, we propagate the PDFs of $\langle\beta_\parallel\rangle$ through equation~\eqref{eqn:map}.
For the purposes of this exercise, we assume that $\sstar/m=4.5\pi\,{\rm cm}^2{\rm g}^{-1}$. 
The expected constraints from clusters at the three redshifts are shown in Figure~\ref{fig:cross_redshift}. 

\begin{figure}
		\begin{centering}
	 		\includegraphics[width = 7cm]{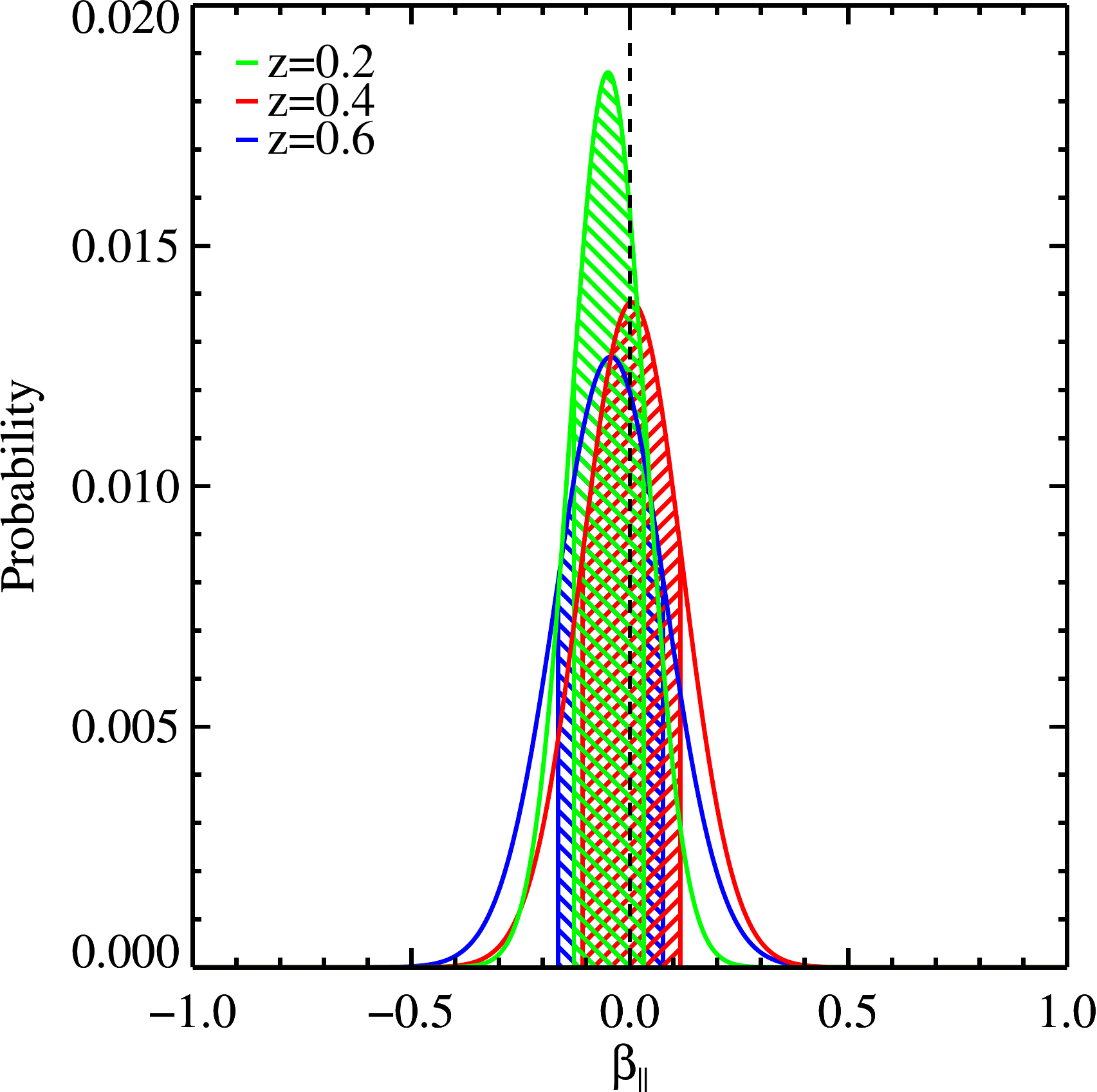}
			\caption{\label{fig:dist_redshift}
			Potential constraints on $\beta_\parallel$ from a sample of 60 minor mergers in the presence of realistic observational noise.
			Hatched regions show the integrated 68\% confidence limits.
			The different colours show expected constraints if the clusters were all at redshift $z_{\rm lens}=0.2$, $0.4$ or $0.6$. 
			We have explicitly removed all redshift-dependence of the physical signal; the changing errors here are due to the apparent angular size and the lensing geometry at different distances from the observer.
			All the distributions are consistent with zero, as expected from the collisionless Dark Matter used in the simulations.}
			\end{centering}
\end{figure}

\begin{figure}
		\begin{centering}
	 		\includegraphics[width = 7cm]{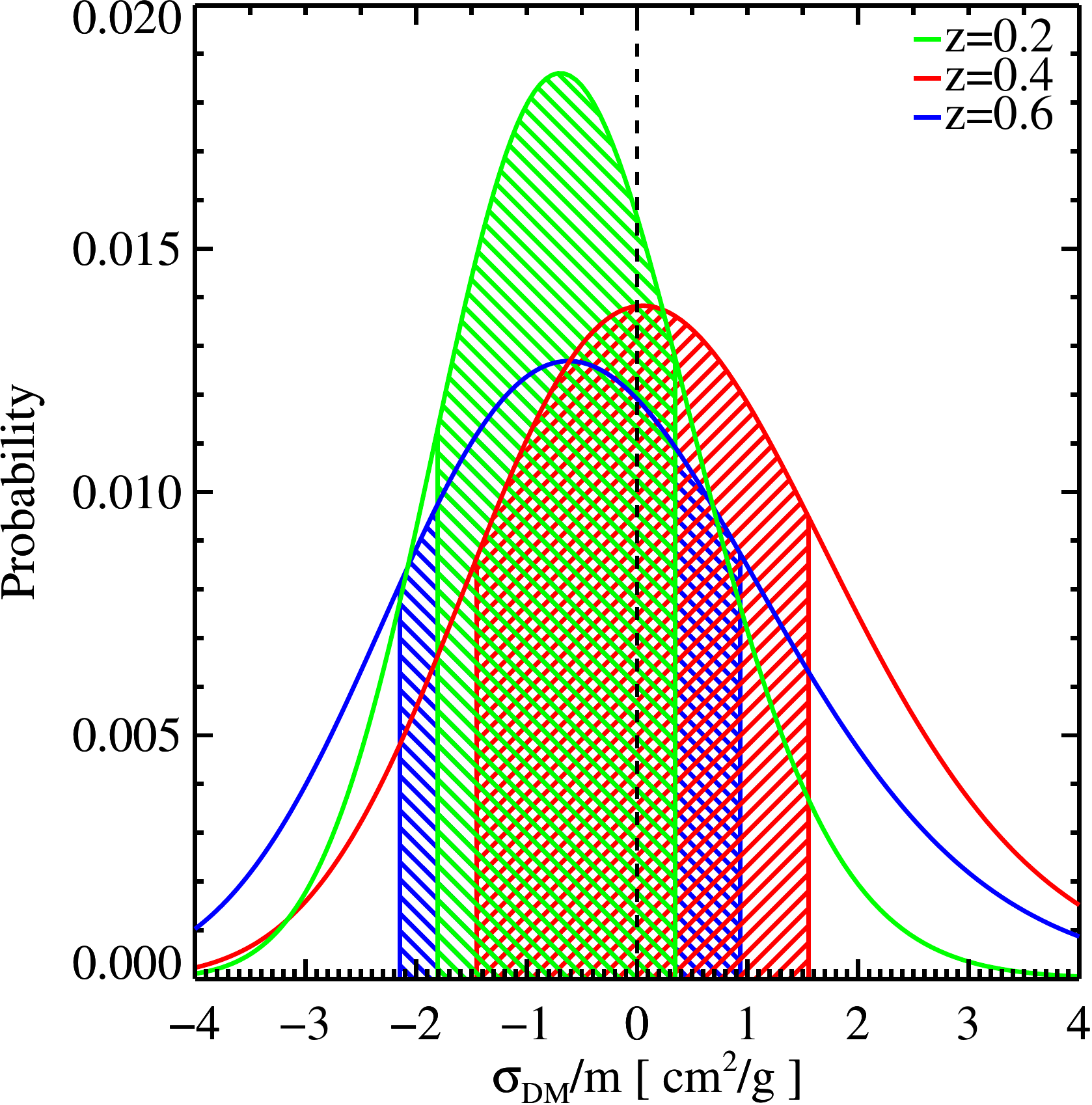}
			\caption{\label{fig:cross_redshift}
			Potential constraints on the self-interaction cross-section of DM $\sdm/\mdm$ from a sample of 60 minor mergers in the presence of realistic observational noise.
			Hatched regions show the integrated 68\% confidence limits.
			This is a propagation of Figure \ref{fig:dist_redshift} using equation \ref{eqn:map}, assuming $\sstar/m=4.5\pi\,{\rm cm}^2{\rm g}^{-1}$. 
			For clusters at redshift $z=0.2$ the constraints are much tighter inferring potentially a limit of $<1{\rm cm}^2{\rm g}^{-1}$ to 68\% confidence. }
			\end{centering}
\end{figure}
We find that a conservative sample size of $\sim60$ sub-halos should constrain $\sdm/m$ to less that $1\,{\rm cm}^2/{\rm g}^{-1}$ at the $68\%$ confidence level (or within those errors if the DM really is collisional). 
We note that although the simulations here are of CDM only, and not SIDM, the error bars gained are not expected to alter in the presence of interacting Dark Matter since they reflect the expected scatter when fitting profiles to amorphous halos of dark matter. We are therefore confident that such constraints can be made in the presence of observational data.
%
%
%
%
In the current regime, this is limited only by $\sqrt{n}$ statistics of the number of pieces of observed substructure.
%
This is extremely encouraging for future detections which will have access to orders of magnitude greater numbers of galaxy clusters.

\section{Conclusions}\label{sec:conc}
We have presented 
a new method to probe the interaction cross-section of Dark Matter ($\sdm/m$). 
By measuring the relative distance that a Dark Matter sub-halo lies from its galactic component with respect to the distance the baryonic gas lies from the same galactic component, we have derived a new parameter $\beta$, which is independent of any line of sight projections.
%
In order to interpret this parameter $\beta$ as a cross-section we have developed an approximate analytic model for substructure infall, considering all the major forces acting on the three components.
In particular, we model our DM interactions based on the type of frequent interactions outlined in K13, with particles exchanging small amounts of momentum, resulting in a overall drag force on the halo. 
This regime means that our interpretation is unique in probing types of DM scattering similar to that of Rutherford scattering, in which the differential cross-section is highly anisotropic.

We show, in the limit that the cross-section of DM is small, that the ratio, $\beta$, of the distance between an in-falling Dark Matter (DM) halo and member galaxies and in-falling gas halo and member galaxies scales linearly with the optical depth of the DM-halo.
In the regime that the cross-section becomes comparable with baryonic gas and the halo becomes optically thick we postulate the scaling of $\sdm/m$ to larger values. 
%
We predict that this scaling follows the general equation for the attenuation of momentum from scattering particles through a medium. This interpretation satisfies our conditions that requires this scaling to be linear in the low limit and tend to some value determined by the macroscopic properties of the halo.
We parameterise this transition regime with $\sstar$ and find that $\sstar=\pi s\Vd^2/M\Vd$ where $s$ and $M$ are the size and mass of the DM halo and is analogous to a sub-halo with an optical depth of unity.

The specific improvement of our method over previous work is to use the position of substructure member galaxies to define the direction of infall.
This removes the dominant uncertainties in previous merging cluster estimates of $\sdm/m$, due to the unknown orientation with respect to the line of sight, and the time of infall.
%
%
%
It also defines a preferred direction in each cluster in which to optimally search for a signal (and a perpendicular direction to use as a systematics test).
We have applied our method to hydrodynamical simulations of galaxy cluster formation.
The expected offset of $\sim20h^{-1}\,$kpc is an order of magnitude larger than without the preferred direction, and should be readily observable with existing archival data. 
We find that one should be able to detect an offset between collisionless DM and gas at $\sim6\sigma$, and measure $\sdm/m$ with 68\% confidence limits of $\pm 1.0\,{\rm cm}^2{\rm g}^{-1}$. 

Our analytic model should be sufficient to look for and interpret measurements of $\sdm/m$ from existing archival data.
However, the main benefit of statistically exploiting minor mergers rather than a few major mergers, is that there is an almost limitless number of them all over the sky.
These will be observed in the next decade by surveys such as Euclid$^1$\footnotetext[1]{http://www.euclid-ec.org} \citep{EUCLID} and WFIRST AFTA \citep{WFIRST} which will observe $>1000$ clusters resulting in potential statistical errors of $<0.1{\rm cm}^2{\rm g}^{-1}$. 
In order to understand the physics of substructure in-fall and the separation of mass components, at the level of accuracy required to interpret those data, we will require accurate simulations of minor mergers with DM of varying cross-sections.


\section*{Acknowledgments} 
The authors are pleased to thank Alan Heavens, Catherine Heymans,  and Jim Dunlop for useful conversations and advice.
DH is supported by an STFC studentship. DH and RM acknowledge support from the European Research Council through grant MIRG-CT-208994. RM and TK  are supported by Royal Society URFs. This work was supported in part by the National Science Foundation under Grant No.\ PHYS-1066293 and the hospitality of the Aspen Center for Physics

\bibliographystyle{mn2e}
\bibliography{bibliography}

\begin{thebibliography}{55}
\expandafter\ifx\csname natexlab\endcsname\relax\def\natexlab#1{#1}\fi

\bibitem[{{Agertz} {et~al}\mbox{.}(2007){Agertz}, {Moore}, {Stadel}, {Potter},
  {Miniati}, {Read}, {Mayer}, {Gawryszczak}, {Kravtsov}, {Nordlund}, {Pearce},
  {Quilis}, {Rudd}, {Springel}, {Stone}, {Tasker}, {Teyssier}, {Wadsley}, \&
  {Walder}}]{AMRvSPH}
{Agertz} O. {et~al.}, 2007, \mnras, 380, 963

\bibitem[{{Bartelmann} \& {Schneider}(2001)}]{BS01}
{Bartelmann} M., {Schneider} P., 2001, \physrep, 340, 291

\bibitem[{{Berezhiani}, {Dolgov} \& {Mohapatra}(1996){Berezhiani}, {Dolgov}, \&
  {Mohapatra}}]{mirrordm}
{Berezhiani} Z.~G., {Dolgov} A.~D., {Mohapatra} R.~N., 1996, Physics Letters B,
  375, 26

\bibitem[{{Booth} \& {Schaye}(2009)}]{AGNfeedback}
{Booth} C.~M., {Schaye} J., 2009, \mnras, 398, 53

\bibitem[{{Brada{\v c}} {et~al}\mbox{.}(2008){Brada{\v c}}, {Allen}, {Treu},
  {Ebeling}, {Massey}, {Morris}, {von der Linden}, \& {Applegate}}]{minibullet}
{Brada{\v c}} M., {Allen} S.~W., {Treu} T., {Ebeling} H., {Massey} R., {Morris}
  R.~G., {von der Linden} A., {Applegate} D., 2008, \apj, 687, 959

\bibitem[{{Brada{\v c}} {et~al}\mbox{.}(2006){Brada{\v c}}, {Clowe},
  {Gonzalez}, {Marshall}, {Forman}, {Jones}, {Markevitch}, {Randall},
  {Schrabback}, \& {Zaritsky}}]{bulletclusterB}
{Brada{\v c}} M. {et~al.}, 2006, \apj, 652, 937

\bibitem[{{Clowe} {et~al}\mbox{.}(2006){Clowe}, {Brada{\v c}}, {Gonzalez},
  {Markevitch}, {Randall}, {Jones}, \& {Zaritsky}}]{separation}
{Clowe} D., {Brada{\v c}} M., {Gonzalez} A.~H., {Markevitch} M., {Randall}
  S.~W., {Jones} C., {Zaritsky} D., 2006, \apjl, 648, L109

\bibitem[{{Clowe}, {Gonzalez} \& {Markevitch}(2004){Clowe}, {Gonzalez}, \&
  {Markevitch}}]{bulletclusterA}
{Clowe} D., {Gonzalez} A., {Markevitch} M., 2004, \apj, 604, 596

\bibitem[{{Clowe} {et~al}\mbox{.}(2012){Clowe}, {Markevitch}, {Brada{\v c}},
  {Gonzalez}, {Chung}, {Massey}, \& {Zaritsky}}]{A520A}
{Clowe} D., {Markevitch} M., {Brada{\v c}} M., {Gonzalez} A.~H., {Chung} S.~M.,
  {Massey} R., {Zaritsky} D., 2012, \apj, 758, 128

\bibitem[{{Dawson} {et~al}\mbox{.}(2012){Dawson}, {Wittman}, {Jee}, {Gee},
  {Hughes}, {Tyson}, {Schmidt}, {Thorman}, {Brada{\v c}}, {Miyazaki}, {Lemaux},
  {Utsumi}, \& {Margoniner}}]{musket}
{Dawson} W.~A. {et~al.}, 2012, \apjl, 747, L42

\bibitem[{{Dubinski} \& {Carlberg}(1991)}]{corecusp}
{Dubinski} J., {Carlberg} R.~G., 1991, \apj, 378, 496

\bibitem[{{Felten} {et~al}\mbox{.}(1966){Felten}, {Gould}, {Stein}, \&
  {Woolf}}]{bremm}
{Felten} J.~E., {Gould} R.~J., {Stein} W.~A., {Woolf} N.~J., 1966, \apj, 146,
  955

\bibitem[{{Feng} {et~al}\mbox{.}(2009){Feng}, {Kaplinghat}, {Tu}, \&
  {Yu}}]{ChargedDM}
{Feng} J.~L., {Kaplinghat} M., {Tu} H., {Yu} H.-B., 2009, Journal of Cosmology
  and Astroparticle Physics, 7, 4

\bibitem[{{Firmani} {et~al}\mbox{.}(2000){Firmani}, {D'Onghia}, {Avila-Reese},
  {Chincarini}, \& {Hern{\'a}ndez}}]{GalaxySIDM}
{Firmani} C., {D'Onghia} E., {Avila-Reese} V., {Chincarini} G., {Hern{\'a}ndez}
  X., 2000, \mnras, 315, L29

\bibitem[{{Frisch}(1995)}]{drageq}
{Frisch} U., 1995, {Turbulence. The legacy of A. N. Kolmogorov.} {Cambridge
  University Press}

\bibitem[{{Fruscione} {et~al}\mbox{.}(2006){Fruscione}, {McDowell}, {Allen},
  {Brickhouse}, {Burke}, {Davis}, {Durham}, {Elvis}, {Galle}, {Harris},
  {Huenemoerder}, {Houck}, {Ishibashi}, {Karovska}, {Nicastro}, {Noble},
  {Nowak}, {Primini}, {Siemiginowska}, {Smith}, \& {Wise}}]{CIAOWavdetect}
{Fruscione} A. {et~al.}, 2006, in Society of Photo-Optical Instrumentation
  Engineers (SPIE) Conference Series, Vol. 6270, Society of Photo-Optical
  Instrumentation Engineers (SPIE) Conference Series

\bibitem[{{Gnedin} \& {Ostriker}(2001)}]{EllGalSIDM}
{Gnedin} O.~Y., {Ostriker} J.~P., 2001, \apj, 561, 61

\bibitem[{{Harvey} {et~al}\mbox{.}(2013){Harvey}, {Massey}, {Kitching},
  {Taylor}, {Jullo}, {Kneib}, {Tittley}, \& {Marshall}}]{Harvey13}
{Harvey} D., {Massey} R., {Kitching} T., {Taylor} A., {Jullo} E., {Kneib}
  J.-P., {Tittley} E., {Marshall} P.~J., 2013, ArXiv e-prints

\bibitem[{{Hoekstra} \& {Jain}(2008)}]{HoekstraRev}
{Hoekstra} H., {Jain} B., 2008, Annual Review of Nuclear and Particle Science,
  58, 99

\bibitem[{{Jee} {et~al}\mbox{.}(2012){Jee}, {Mahdavi}, {Hoekstra}, {Babul},
  {Dalcanton}, {Carroll}, \& {Capak}}]{A520B}
{Jee} M.~J., {Mahdavi} A., {Hoekstra} H., {Babul} A.~., {Dalcanton} J.~J.,
  {Carroll} P., {Capak} P., 2012, \apj, 747, 96

\bibitem[{{Jullo} {et~al}\mbox{.}(2007){Jullo}, {Kneib}, {Limousin},
  {El{\'{\i}}asd{\'o}ttir}, {Marshall}, \& {Verdugo}}]{lenstool}
{Jullo} E., {Kneib} J.-P., {Limousin} M., {El{\'{\i}}asd{\'o}ttir} {\'A}.,
  {Marshall} P.~J., {Verdugo} T., 2007, New Journal of Physics, 9, 447

\bibitem[{{Kahlhoefer} {et~al}\mbox{.}(2013){Kahlhoefer}, {Schmidt-Hoberg},
  {Frandsen}, \& {Sarkar}}]{SIDMModel}
{Kahlhoefer} F., {Schmidt-Hoberg} K., {Frandsen} M.~T., {Sarkar} S., 2013,
  ArXiv e-prints

\bibitem[{{Kaiser} \& {Squires}(1993)}]{KS93}
{Kaiser} N., {Squires} G., 1993, \apj, 404, 441

\bibitem[{{Kauffmann}, {White} \& {Guiderdoni}(1993){Kauffmann}, {White}, \&
  {Guiderdoni}}]{satelliteprob}
{Kauffmann} G., {White} S.~D.~M., {Guiderdoni} B., 1993, \mnras, 264, 201

\bibitem[{{Larson} {et~al}\mbox{.}(2011){Larson}, {Dunkley}, {Hinshaw},
  {Komatsu}, {Nolta}, {Bennett}, {Gold}, {Halpern}, {Hill}, {Jarosik}, {Kogut},
  {Limon}, {Meyer}, {Odegard}, {Page}, {Smith}, {Spergel}, {Tucker}, {Weiland},
  {Wollack}, \& {Wright}}]{WMAP7}
{Larson} D. {et~al.}, 2011, \apjs, 192, 16

\bibitem[{{Laureijs} {et~al}\mbox{.}(2011){Laureijs}, {Amiaux}, {Arduini},
  {Augu{\`e}res}, {Brinchmann}, {Cole}, {Cropper}, {Dabin}, {Duvet}, {Ealet},
  \& et~al.}]{EUCLID}
{Laureijs} R. {et~al.}, 2011, ArXiv e-prints

\bibitem[{{Leauthaud} {et~al}\mbox{.}(2007){Leauthaud}, {Massey}, {Kneib},
  {Rhodes}, {Johnston}, {Capak}, {Heymans}, {Ellis}, {Koekemoer}, {Le
  F{\`e}vre}, {Mellier}, {R{\'e}fr{\'e}gier}, {Robin}, {Scoville}, {Tasca},
  {Taylor}, \& {Van Waerbeke}}]{COSMOSintdisp}
{Leauthaud} A. {et~al.}, 2007, \apjs, 172, 219

\bibitem[{{Loeb} \& {Weiner}(2011)}]{DMYukawa}
{Loeb} A., {Weiner} N., 2011, Physical Review Letters, 106, 171302

\bibitem[{{Mahdavi} {et~al}\mbox{.}(2007){Mahdavi}, {Hoekstra}, {Babul},
  {Balam}, \& {Capak}}]{A520}
{Mahdavi} A., {Hoekstra} H., {Babul} A., {Balam} D.~D., {Capak} P.~L., 2007,
  \apj, 668, 806

\bibitem[{{Markevitch} {et~al}\mbox{.}(2004){Markevitch}, {Gonzalez}, {Clowe},
  {Vikhlinin}, {Forman}, {Jones}, {Murray}, \& {Tucker}}]{bulletcluster}
{Markevitch} M., {Gonzalez} A.~H., {Clowe} D., {Vikhlinin} A., {Forman} W.,
  {Jones} C., {Murray} S., {Tucker} W., 2004, \apj, 606, 819

\bibitem[{{Massey}, {Kitching} \& {Nagai}(2011){Massey}, {Kitching}, \&
  {Nagai}}]{bulleticity}
{Massey} R., {Kitching} T., {Nagai} D., 2011, \mnras, 413, 1709

\bibitem[{{Massey}, {Kitching} \& {Richard}(2010){Massey}, {Kitching}, \&
  {Richard}}]{MKRev}
{Massey} R., {Kitching} T., {Richard} J., 2010, Reports on Progress in Physics,
  73, 086901

\bibitem[{{Meneghetti} {et~al}\mbox{.}(2001){Meneghetti}, {Yoshida},
  {Bartelmann}, {Moscardini}, {Springel}, {Tormen}, \& {White}}]{SIDMCore}
{Meneghetti} M., {Yoshida} N., {Bartelmann} M., {Moscardini} L., {Springel} V.,
  {Tormen} G., {White} S.~D.~M., 2001, \mnras, 325, 435

\bibitem[{{Merten} {et~al}\mbox{.}(2011){Merten}, {Coe}, {Dupke}, {Massey},
  {Zitrin}, {Cypriano}, {Okabe}, {Frye}, {Braglia}, {Jim{\'e}nez-Teja},
  {Ben{\'{\i}}tez}, {Broadhurst}, {Rhodes}, {Meneghetti}, {Moustakas},
  {Sodr{\'e}}, {Krick}, \& {Bregman}}]{A2744}
{Merten} J. {et~al.}, 2011, \mnras, 417, 333

\bibitem[{{Miralda-Escud{\'e}}(2002)}]{SIDMTest}
{Miralda-Escud{\'e}} J., 2002, \apj, 564, 60

\bibitem[{{Mohapatra}, {Nussinov} \& {Teplitz}(2002){Mohapatra}, {Nussinov}, \&
  {Teplitz}}]{mirrordm2}
{Mohapatra} R.~N., {Nussinov} S., {Teplitz} V.~L., 2002, \prd, 66, 063002

\bibitem[{{Newton} \& {Kay}(2013)}]{starform}
{Newton} R.~D.~A., {Kay} S.~T., 2013, ArXiv e-prints

\bibitem[{{Peter}(2012)}]{PeterRev}
{Peter} A.~H.~G., 2012, ArXiv e-prints

\bibitem[{{Peter} {et~al}\mbox{.}(2013){Peter}, {Rocha}, {Bullock}, \&
  {Kaplinghat}}]{SIDMSim}
{Peter} A.~H.~G., {Rocha} M., {Bullock} J.~S., {Kaplinghat} M., 2013, \mnras,
  430, 105

\bibitem[{{Planck Collaboration} {et~al}\mbox{.}(2013){Planck Collaboration},
  {Ade}, {Aghanim}, {Armitage-Caplan}, {Arnaud}, {Ashdown}, {Atrio-Barandela},
  {Aumont}, {Baccigalupi}, {Banday}, \& et~al.}]{planckpars}
{Planck Collaboration} {et~al.}, 2013, ArXiv e-prints

\bibitem[{{Pospelov}, {Ritz} \& {Voloshin}(2008){Pospelov}, {Ritz}, \&
  {Voloshin}}]{SecludedDM}
{Pospelov} M., {Ritz} A., {Voloshin} M., 2008, Physics Letters B, 662, 53

\bibitem[{{Powell}, {Kay} \& {Babul}(2009){Powell}, {Kay}, \&
  {Babul}}]{Powell09}
{Powell} L.~C., {Kay} S.~T., {Babul} A., 2009, \mnras, 400, 705

\bibitem[{{Randall} {et~al}\mbox{.}(2009){Randall}, {Jones}, {Markevitch},
  {Blanton}, {Nulsen}, \& {Forman}}]{impactpars}
{Randall} S.~W., {Jones} C., {Markevitch} M., {Blanton} E.~L., {Nulsen}
  P.~E.~J., {Forman} W.~R., 2009, \apj, 700, 1404

\bibitem[{{Refregier}(2003)}]{RefregierRev}
{Refregier} A., 2003, \araa, 41, 645

\bibitem[{{Rocha} {et~al}\mbox{.}(2013){Rocha}, {Peter}, {Bullock},
  {Kaplinghat}, {Garrison-Kimmel}, {O{\~n}orbe}, \& {Moustakas}}]{SIDMSimA}
{Rocha} M., {Peter} A.~H.~G., {Bullock} J.~S., {Kaplinghat} M.,
  {Garrison-Kimmel} S., {O{\~n}orbe} J., {Moustakas} L.~A., 2013, \mnras, 430,
  81

\bibitem[{{Shan}, {Qin} \& {Zhao}(2010){Shan}, {Qin}, \& {Zhao}}]{rareevents}
{Shan} H.~Y., {Qin} B., {Zhao} H.~S., 2010, \mnras, 408, 1277

\bibitem[{{Spergel} {et~al}\mbox{.}(2013){Spergel}, {Gehrels}, {Breckinridge},
  {Donahue}, {Dressler}, {Gaudi}, {Greene}, {Guyon}, {Hirata}, {Kalirai},
  {Kasdin}, {Moos}, {Perlmutter}, {Postman}, {Rauscher}, {Rhodes}, {Wang},
  {Weinberg}, {Centrella}, {Traub}, {Baltay}, {Colbert}, {Bennett},
  {Kiessling}, {Macintosh}, {Merten}, {Mortonson}, {Penny}, {Rozo},
  {Savransky}, {Stapelfeldt}, {Zu}, {Baker}, {Cheng}, {Content}, {Dooley},
  {Foote}, {Goullioud}, {Grady}, {Jackson}, {Kruk}, {Levine}, {Melton},
  {Peddie}, {Ruffa}, \& {Shaklan}}]{WFIRST}
{Spergel} D. {et~al.}, 2013, ArXiv e-prints

\bibitem[{{Spergel} \& {Steinhardt}(2000)}]{ObserveSIDM}
{Spergel} D.~N., {Steinhardt} P.~J., 2000, Physical Review Letters, 84, 3760

\bibitem[{{Springel}(2005)}]{gadget2}
{Springel} V., 2005, \mnras, 364, 1105

\bibitem[{{Springel} {et~al}\mbox{.}(2008){Springel}, {Wang}, {Vogelsberger},
  {Ludlow}, {Jenkins}, {Helmi}, {Navarro}, {Frenk}, \& {White}}]{CDMsatellites}
{Springel} V. {et~al.}, 2008, \mnras, 391, 1685

\bibitem[{{Thacker} {et~al}\mbox{.}(2000){Thacker}, {Tittley}, {Pearce},
  {Couchman}, \& {Thomas}}]{turbulence}
{Thacker} R.~J., {Tittley} E.~R., {Pearce} F.~R., {Couchman} H.~M.~P., {Thomas}
  P.~A., 2000, \mnras, 319, 619

\bibitem[{{Vogelsberger}, {Zavala} \& {Loeb}(2012){Vogelsberger}, {Zavala}, \&
  {Loeb}}]{SubhalosSIDM}
{Vogelsberger} M., {Zavala} J., {Loeb} A., 2012, \mnras, 423, 3740

\bibitem[{{Watson} {et~al}\mbox{.}(2013){Watson}, {Iliev}, {Diego},
  {Gottl{\"o}ber}, {Knebe}, {Mart{\'{\i}}nez-Gonz{\'a}lez}, \&
  {Yepes}}]{extremeObjects}
{Watson} W.~A., {Iliev} I.~T., {Diego} J.~M., {Gottl{\"o}ber} S., {Knebe} A.,
  {Mart{\'{\i}}nez-Gonz{\'a}lez} E., {Yepes} G., 2013, ArXiv e-prints

\bibitem[{{Williams} \& {Saha}(2011)}]{cannibal}
{Williams} L.~L.~R., {Saha} P., 2011, \mnras, 415, 448

\bibitem[{{Yoshida} {et~al}\mbox{.}(2000){Yoshida}, {Springel}, {White}, \&
  {Tormen}}]{HaloSIDM}
{Yoshida} N., {Springel} V., {White} S.~D.~M., {Tormen} G., 2000, \apjl, 544,
  L87

\end{thebibliography}

\appendix

\bsp
\label{lastpage}

\end{document}